\documentclass{aa}
\usepackage{graphics}

\def\etal {{\rm et al.~}} 

\def\ltsima{$\; \buildrel < \over \sim \;$}
\def\simlt{\lower.5ex\hbox{\ltsima}}
\def\gtsima{$\; \buildrel > \over \sim \;$}
\def\simgt{\lower.5ex\hbox{\gtsima}}

\def\vs{{~vs.~}}

\def\kms{\ifmmode {\rm \ km \ s^{-1}}  
\else
$\rm km \ s^{-1}$\fi}

\def\h2{\mbox{\ion{H}{ii}}}

\newcommand{\lessim}{\mathrel{\hbox{\rlap{\hbox{\lower4pt\hbox{$\sim$}}}\hbox{$<$}}}}

\begin{document}
\addtolength{\voffset}{1cm}

\thesaurus{08.12.3; 11.03.1; 11.13.1}

\title{Stellar luminosity functions of rich star clusters in the Large Magellanic Cloud}

\author{Bas\'\i lio Santiago$^1$, Sylvie Beaulieu$^2$, Rachel Johnson$^2$, Gerard F. Gilmore$^2$ }

\offprints{Bas\'\i lio X. Santiago}

\institute{Universidade Federal Rio Grande do Sul, IF, 
CP\,15051, Porto Alegre 91501-970, RS, Brasil
\and
Institute of Astronomy, Madingley Rd.,Cambridge CB3 0HA, UK}

\date{Received; accepted}

\titlerunning{Stellar Luminosity Functions of Rich Star Clusters in the LMC}
\authorrunning{Santiago \etal}

\maketitle

\begin{abstract}
We show the results of deep $V$ and $I$ HST photometry of 6 rich star 
clusters in the Large Magellanic Cloud with different ages and 
metallicities. The number of stars with measured magnitudes
in each cluster varies from about 3000 to 10000 stars.
We build stellar density and surface brightness profiles for the clusters
and extract half-light radii and other
structural parameters for each. We also
obtain luminosity functions, $\Phi (M_V)$, down to $M_V \simeq 6$ 
($M/M_{\odot}$ \gtsima $0.9$), and investigate their dependence with
distance from the cluster centre 
well beyond their half-light radius. In all clusters we find 
a systematic increase in the
luminosity functions slope, $\Delta log \Phi (M_V) / \Delta (M_V)$, with radial
distance from the centre. Among the clusters displaying significant 
mass segregation are the 2 youngest in the sample: NGC 1805, NGC 1818.
For these two clusters we obtain present day mass functions. 
The NGC 1818 mass function is in excellent agreement with that derived 
by other authors, also using HST data. The young cluster mass function slopes differ, 
that of NGC 1805 being systematically steeper than NGC 1818. 
Since these are very young stellar systems ($\tau$ \ltsima $40$ Myrs),
these variations may reflect the initial conditions rather than
evolution due to internal dynamics.

\keywords{Stars: luminosity function, mass function; Galaxies: clusters: general; Magellanic Clouds}

\end{abstract}

\section{Introduction}
\label{intro}
Rich star clusters are made up of tens and sometimes hundreds of
thousands of stars bound together by gravity. Models for the formation
and evolution of such systems have to incorporate effects of dynamical
friction, stellar evaporation, tidal effects, disc shocks, two-body
interactions and stellar evolution (Heggie \& Aarseth 1992, Spurzem 
\& Aarseth 1996, Vesperini \& Heggie 1997, Goodwin 1997a,
Aarseth 1999).  Substantial progress is being made towards
understanding such dynamical effects and realistically modeling them.
Model predictions generally make use of N-body simulations and can
more easily be tested for a particular cluster (or set of clusters) if
enough information about its present day structure, shape, density and
velocity field is available for comparison (Vesperini \& Heggie 1997,
Goodwin 1997b, de Oliveira \etal 2000).  With N-body models one can
also predict the luminosity function and mass function for a
cluster at some distance from the centre and at some age. This process
yields insight on the initial conditions under which the cluster(s)
formed, i.e., the initial mass function (IMF), density profile, amount
of original stellar segregation, etc. The stellar IMF, in particular,
is a very valuable piece of information as many fundamental
astrophysical issues depend on its shape and possible universality
(Chiosi \etal 1998).

Recovering initial conditions from the current properties of a cluster
becomes harder the older it is, since more time has elapsed for
internal dynamics and stellar evolution to erase them.  Unfortunately,
our Milky-Way galaxy contains a system of purely old (ages $\tau$
\gtsima $10^{10}$ years) rich globular clusters, mostly located in the
Galactic halo.  The nearest galaxy with a diverse system of rich star
clusters, spanning a wide range in ages, metallicities and positions
is the Large Magellanic Cloud (LMC) (Westerlund 1990).  It is
therefore a suitable laboratory for studying the processes by which
clusters form and evolve.

In this paper we describe the analysis of density profiles and
luminosity functions of 6 rich LMC clusters, deliberately chosen so as
to encompass a large range in age and metallicity. The results are
intended as a useful observational input to modelling by N-body
experiments (including those we have underway). The data come from
Hubble Space Telescope (HST) observations using the Wide Field and
Planetary Camera 2 (WFPC2). With the high-resolution imaging provided
by this instrumental setup it was possible to obtain luminosity functions deeper than
in any previous attempt.  The paper is outlined as follows: in Sect. 2 we
describe the data, the stellar samples derived from them, and the
photometry. Density profiles and position-dependent luminosity functions are presented
in Sect. 3. In Sect. 4 we discuss our results and compare them with other
works. Finally, in Sect. 5 we present our conclusions.

\section{Observations}

These WFPC2 data are part of the cycle 7 HST 7307 project (Beaulieu
\etal 1998, Elson \etal 1998a,b, Johnson \etal 1998). The project
includes observations of 8 clusters: NGC 1805, NGC 1818, NGC 1831, NGC
1868, NGC 2209, Hodge 14, Hodge 11 and NGC 2210. The two latter are
old systems ($\tau > 10^{10}$ years) and therefore have a much fainter
main-sequence turn-off than the others. As our main interest is to
analyze the shape of main-sequence luminosity functions, we here concentrate on the
remaining 6 clusters.

For each cluster, two sets of WFPC2 images were obtained, one
with the centre of the Planetary Camera (PC) coinciding with the
cluster's center and the other with the PC centered on the the
cluster's half-light radius.  We call these the CEN and HALF fields,
respectively.  6 exposures were obtained with each of the F814W and
F555W filters in the CEN field: 3 exposures were deeper (individual
exposure times of 300s and 140s, respectively, for F814W and F555W),
the remaining 3 were shallow (20s and 5s, respectively, for F814W and
F555W); the shallower exposures were meant to allow for photometry of
bright stars.  For the HALF fields, 3 deep exposures (total exposure
time: 2500s) were taken with each filter.  Saturation for these latter
was a problem only for $m_{555}$ \ltsima $19$.

WFPC2 images located about 7' away from each cluster are also
available as part of the 7307 project. These are parallel observations
with NICMOS as the primary instrument. Two exposures were taken with
each of the two filters, the total exposure time being 1200s for F555W
and 800s for F814W. These were used for subtraction of background
field LMC stars.  Colour-magnitude diagrams (CMDs) for the field stars
in the images were studied by Castro \etal (2000).

The 3 individual exposures at each pointing, filter and exposure time
were combined together. No registration was required.  The same was
done for the 2 exposures in the parallel fields used for field star
subtraction.  The final stacked image in each case was then used for
sample selection and photometry.

In Table 1 we list the basic properties of the clusters plus the total
number of stars found in our images. The data in the table were taken
from Elson (1991), Corsi \etal (1994), Will \etal (1995), Mateo (1988) and
Johnson \etal (2000).

\begin{table}
\caption[]{Cluster's parameters.}
\label{tab1}
\small
\renewcommand{\tabcolsep}{1.1mm}
\begin{tabular}{llllll}
\hline\hline
Cluster & $\theta_{LMC} (\circ)$ & E(B-V) & log Age (yrs) & $[Fe/H]$ & $N_{stars}$ \\ \hline
\hline
NGC 1805 & 4.1 & 0.04  & 7.0-7.3 & $-0.4$ & 6508 \cr
NGC 1818 & 3.8 & 0.03  & 7.3-7.6 & $-0.4$ & 10187 \cr
NGC 1831 & 5.1 & 0.05  & 8.5-8.7 & $-0.3$ & 7801 \cr
NGC 1868 & 5.9 & 0.02  & 8.8 & $-0.6$ & 7362 \cr
NGC 2209 & 5.4 & 0.07  & 8.9-9.1 & $-1.0$ & 3771 \cr
Hodge 14 & 3.9 & 0.04  & 9.2-9.4 & $-0.7$ & 3133 \cr
\hline\hline
\end{tabular}
\end{table}

\subsection{Sample Selection}

The IRAF DAOPHOT package was used to detect and classify sources
automatically in each stacked F814W image.  Star candidates were
detected using DAOFIND, with a peak intensity threshold for detection
set to $5~\sigma$, where $\sigma$ corresponds to the $rms$ fluctuation
in the sky counts, determined individually for each image.
 
A preliminary aperture photometry was carried out by running the task
PHOT on all objects output by DAOFIND. Star/galaxy separation required
modelling point sources by fitting the profiles of bright, isolated
and non-saturated stars to a moffat function ($\beta = 1.5$). We used
the PSF task for that purpose.  This PSF model was then fitted to all
objects found by DAOFIND with the ALLSTAR task.

The output parameters given by ALLSTAR (fit quality $\chi$, magnitude
uncertainty $\delta m$, and sharpness $s$) are useful to separate a
purely stellar sample from the remaining sources detected. An
effective use of these parameters was made by finding the linear
combination of $\chi$, $\delta m$ and $s$ that best separated a sample
of eye-ball selected stars from a sample of eye-ball non-stellar
sources (either galaxies or spurious detections around bright
saturated stars and CCD defects).  These eye-ball samples were
selected on each chip of each image and the best linear combination
was found by varying the linear coefficients and cut-off value until
they led to a complete (or nearly) separation between the two samples
in parameter space.  This star/galaxy classifier was then applied to
all objects in the list of candidates and yielded samples of
stars. The final stellar samples selected were inspected on the images
in order to check if star/galaxy separation was effective and that
obvious stars were not wrongly discarded.  The whole procedure was
repeated if necessary until a satisfactory star sample selection was
achieved.

This sample selection procedure was applied to both cluster (CEN and
HALF fields) and background images.

\subsection{Photometry}

Due to the undersampling of point source light distributions with the
WFPC2, aperture photometry tends to yield more accurate magnitude and
color measurements than psf fitting.  Therefore, aperture photometry
was carried out on the stars selected in each image, on both the
$F814W$ and $F555W$ images. A 2 pixel aperture radius was used for
all WFPC2 images and for both filters, corresponding to r = 0.2''
for the 3 Wide Field Camera (WFC) chips and r = 0.09'' for the Planetary
Camera (PC); this is a compromise
between the need to include the core of the psf but avoid light
contamination by neighbouring objects.

Significant but well determined aperture corrections were then applied
to the measured magnitudes (Holtzman \etal 1995a).  These corrections
are larger for the PC chip than for the 3 WFC ones, being about the
same on the F814W and F555W filters.  All the data have been
zero-pointed, CTE and aperture corrected following the prescriptions
of Holtzman \etal (1995a,b) and Whitmore \etal (1999). The CTE
corrections from the latter were used only for the shallow images in
the CEN fields.  For the deep exposures, the difference between the
Holtzman \etal ramp correction and the Whitmore \etal time-dependent
one is very small.

A single reddening correction was adopted for the clusters CEN and HALF fields
and for their corresponding background field.
The E(B-V) adopted values are listed in Table 1.

One advantage of having the partially overlapping CEN and HALF images
is the ability to estimate the accuracy of our aperture photometry and
to search for any systematic error in the measured magnitudes and
colors.  The stars in the two fields were matched by their equatorial
coordinates, as provided by the astrometric solution available in the
image headers.  These solutions are internally consistent but often
have small ($\simeq 3 - 5 ~pixels$) offsets from one image to
another. These offsets were measured by eye by superposing the
estimated positions of the CEN and HALF stars. Only stars in the HALF
and CEN fields with corrected positions within 1 WFC pixel of each
other were considered successful matches.

The photometric comparison for the stars in common between the HALF
and CEN fields of NGC 1805 is shown in Fig. 1. The upper panel shows
the $m_{814}$ (solid squares) and $m_{555}$ (open circles)
values. Except for the effects of saturation at the bright end in the
magnitude comparison, the data lie very close to the identity lines in
both cases, indicating the lack of any significant systematics in the
photometry. A slight systematic effect is seen in F814W as shown by
the linear fit to the data shown. This best fit solution,

$$m_{CEN} = 0.98~m_{HALF} + 0.40,$$

was used to bring the magnitudes of the HALF field to the same system as the
CEN ones.

The few outliers in Fig. 1 were investigated further and proved to be
positional $CEN \vs HALF$ mismatches that escaped our positional matching
algorithm. Fig. 1 is typical of all the other clusters.

\subsection{Completeness Corrections}

The non-detection of faint stars by the automatic detection and star selection
algorithms was quantified by simulations of artificial stars.
For each field, pointing and chip,
a total of 1000 artificial stars of a
fixed magnitude were added to the F814W images.
These 1000 stars were placed in random positions on the image,
in 10 realizations with 100 stars each.
Magnitude bins of $0.5$ mag width in the range 
$20 \leq m_{F814} \leq 26$ were covered by the experiments. 
Since the completeness correction depends
on crowding, completeness was computed as a function of position
on each CCD chip; the fraction of artificial stars
recovered in each one of 64 sectors of $100\times100$ pixels in each chip
was taken as the
completeness value $c$ for that sector. To correct for incompleteness, each 
star was then given a weight $w$ corresponding to the distance-weighted
average of $1/c$ values of the 4 closest sectors to it:

$$w_i = { {1} \over  {\sum_j r_j^{-1}} }~~\sum_{j}  { {r_j^{-1}} \over {c (m_i,f,k,s_j)} }, \eqno (1)$$

where $i$ is an index for the star (with apparent magnitude $m_i$), 
$c$ is the completeness value at
its field $f$, chip $k$ and at sector $s_j$, whose centre lies at
a distance $r_j$ from the star position on the chip, 
$j$ being an index for the 4
sectors which are closest to the star.

Fig. 2 shows an example case of completeness values for
the PC1 chip of the NGC 1868 CEN field. The automatically detected objects
in the chip are shown as dots in the figure as a function 
of X and Y positions. 
The completeness values for $23.0 \leq m_{814} \leq 23.5$
found by the procedure described above
are also shown for each sector on the chip. There is a clear
anticorrelation between crowding and completeness, as expected: in the central
cluster regions completeness is significantly smaller than in the outer areas.

Completeness functions were also computed for the parallel 
WFPC2 fields used for field star subtraction. 
The low stellar density of these fields allowed us to use a
single completeness value for each magnitude bin.

\section{Density Profiles and Luminosity Functions}

In order to build density or surface brightness profiles and to 
search for variations in the shape of the luminosity
function with position within each cluster, the stars
found in the HALF and CEN fields 
had to be put together into a single sample. This combined sample
included all the stars found in the CEN field plus the stars in the
HALF field located outside the CEN field borders.
We again used the astrometric information available in the image
headers (plus any additional offset correction; see Sect. 2.2) 
to find the equatorial coordinates (2000 EPOCH)
corresponding to the CEN field chip borders.

Fig. 3 shows the distribution on the sky
of the combined sample of stars for NGC 1805. 
The extra stars added by the HALF field
are shown with different symbols for clarity.

An important factor to be accounted for when computing the profiles
and luminosity functions for each cluster is the increasing fractional 
loss of sampled area in the sky
as a function of angular distance from the cluster's center. 
The profiles and luminosity functions
were considered only out to radii where the  
area sampled by the combined
CEN and HALF fields was larger than 30\% of the total for that radius.
This typically corresponded to 
$R$ \ltsima $25$ pc (roughly 1000 pixels). Within these limits,
each star in our sample was given an additional
weight corresponding to the inverse of the fractional
sampled area at its distance
from the cluster centre. This fraction was computed 
by randomly throwing points within each ring of a fixed radius and determining
the percentage of such points within either the CEN or HALF field borders.
Therefore, the final weight assigned to each star was:

$$W_i = { {w_i} \over {f(r_i)} }, \eqno (2)$$

where $i$ is an index for the stars, $w_i$ is given by eq. (1), and $f(r_i)$
is simply

$${ {A_{samp} (r_i)} \over {A_{tot} (r_i)} }, \eqno (3)$$

where $A_{samp} (r_i)$ is the sampled area (CEN or HALF fields added together)
and $A_{tot} = 2~\pi~r_i~dr_i$ is the total area of a ring of width $dr_i$
at a distance $r_i$ from the cluster centre.

In the next subsections we show density and surface brightness profiles
and luminosity functions for each cluster in the sample.

\subsection{NGC 1805}

Fig. 4 shows the number density and surface brightness profiles of NGC 1805
out to the largest possible radii available. 
The profiles were built by adding up the weighted number of stars in 
rings 150 pixels ($\simeq 3.6$ pc) wide centered on the cluster. 
Each star contributed to the 
profile with a weight given by eqs (1) and (2) as explained in the previous 
section.

The upper curves in Fig. 4
are not corrected for field star contamination. They roughly
flatten at $R \simeq 10$ pc $\simeq 410$ pixels from the cluster's center. 
The lower curves show the cluster profiles after the contribution of field
stars in the background was subtracted off. 

From its light profile we were able to estimate the half-light radius
of NGC 1805 to be $R_{hl} \simeq 1.8$ pc.

In Fig. 5 we show the distribution of stars as a function of
absolute magnitude $M_{555}$ ($\Phi (M_{555})$ or luminosity function)
for different regions of the cluster.  The $M_{555}$ values assume a
distance modulus of $m - M = 18.5$ for the LMC (Panagia \etal
1991). Only main-sequence stars were considered and the field star
contamination was subtracted off using the background WFPC2
field. Even though luminosity functions could extend brightwards in the CEN field, we
did not analyze the luminosity function bright end because saturation sets in at
$M_{555} \simeq 18.5-19$ for stars in the HALF field, therefore
considerably reducing the available solid angle.  All the luminosity functions were
scaled to the total cluster area used to build the global luminosity function, shown in
panel {\it a} of each figure.

Panel 5a shows the global luminosity function within $R = 1050$ pix 
($\simeq 25$ pc).  The other panels probe different distances from the centre,
as indicated.  The luminosity function in the innermost radius is clearly flattened
compared to the ones further out, there being in fact a systematic
trend of increase in the luminosity function slopes with distance from the cluster
centre. Linear fits were carried out for each luminosity function and are also shown in
Fig. 5. With the exception of the last radial ring, a power-law in
luminosity is a good description of the luminosity function shapes.

The best-fit slope values are plotted in Fig. 6. The systematic
increase in the slopes is clear out to about 15 pc from the
center. There is a drop in the luminosity function slope in the 
last radial rings, but
the density contrast above the background is small and therefore the
statistics are noisier.  Given the young age of this cluster (Johnson
\etal 2000), the observed mass segregation in NGC 1805 possibly
reflects the initial conditions during cluster formation (see Sect. 4).

\subsection{NGC 1818}

Fig. 7 shows the stellar density and surface brightness profiles
of NGC 1818. The symbols are the same as in Fig. 4.
NGC 1818 is more extended  ($R_{hl} = 2.6$ pc) and has
a steeper profile than NGC 1805 in the outer regions; 
its corrected profiles systematically decline out to the 
outermost radius shown. At $r \simeq 18$ pc
the cluster density has fallen by about a factor of 50 
from its central value of $\sigma(0) \simeq 40$ stars/pc$^{2}$.

Luminosity functions for various rings (again each 3.6 pc wide) 
centred on NGC 1818 are
shown in Fig. 8.  As in NGC 1805, the luminosity function slopes tend to
systematically increase with radius out to $R \simeq 17$ pc from the
centre (Fig. 9).  A drop in the luminosity function slope is seen in the outermost
radius of both clusters.  The observed positional dependence of the luminosity function
slopes may again be primordial, since NGC 1818 is also a young system
(Johnson \etal 2000).

\subsection{NGC 1831}

The density and surface brightness profiles of NGC 1831 are shown in
Fig. 10. The number density profile is steeper than that of NGC 1818
but the surface brightness profile is shallower. NGC 1831 is richer
and has a higher central density of stars ($\sigma \simeq
50$ stars/pc$^{2}$) than either NGC 1805 or NGC 1818, but its central
surface brightness is lower than that of NGC 1818.  We obtain a
half-light radius of $R_{hl} \simeq 3.6$ pc from the light
distribution of NGC 1831.

The positional dependent luminosity functions are shown in Fig. 11.  Comparison of
the luminosity functions in panel 11b with those in panel 11c shows that
the inner 7 or 8 pc ($R$ \ltsima $300$ pix) contains a larger relative
proportion of luminous ($M_V$ \ltsima $2$) stars.  Notice that the luminosity function
in the innermost radial bin shows an abrupt drop at $M_{555} >
5.5$. This could be an artificiality caused by an overestimated
completeness function. A less abrupt drop is present in the central luminosity function
of NGC 1818 as well. A straight line in log $\Phi$ \vs $M_{555}$ seems
to be a better description of the luminosity function shapes in NGC 1831 than it was
for NGC 1818.

Mass segregation is also confirmed in Fig. 12, where the best-fit
slopes to the luminosity functions are plotted; the luminosity function slopes of the 2 inner radii are
smaller than those in the other rings, where the slopes roughly
flatten out.

\subsection{NGC 1868}

This is a richer cluster than the previous ones, with a central
stellar density of about $60$ stars/pc$^{2}$.  NGC 1868 density and
surface brightness profiles extend out nearly to the largest radius
used, as shown by Fig. 13.  It has a steeper profile than NGC 1831,
especially in the inner regions.  In fact, the profiles shown are the
steepest of all clusters in our sample.  From its light profile we
estimate a half-light radius of only about $R_{hl} \simeq 1.8$ pc.

Figs. 14 and 15 show how the luminosity function varies as a function of position
within the cluster. The luminosity function fit in the innermost radial bin is strongly
affected by a drop in the luminosity function at the faintest magnitude bin, which
could again indicate overestimated completeness at this magnitude
level.  But the effect of mass segregation in NGC 1868 is still
clearly visible, with steeper luminosity functions in the outer regions relative to the
inner ones.  The systematic in the slope is clear in Fig. 15 as
well, where the slope continually rises out to $R \simeq 10$ pc,
roughly flatenning out beyond this point.

\subsection{NGC 2209}

NGC 2209 is a considerably sparser cluster than the previous ones. 
It has the lowest central density in both stars and light
($\sigma (0) \simeq 11$ stars/pc$^{2}$; $\mu_V \simeq 21.6$ mag/arcsec$^{2}$),
as attested by Fig. 16. Its half-light radius is large: $R_{hl} \sim 7.6$ pc,
its light profile is therefore the shallowest in our sample.

Fig. 17 shows that a power-law in luminosity is a poor 
description for the luminosity function in NGC 2209.
But the linear fits
and derived slopes, despite not always reflecting the complexity in the luminosity function
shapes, can be still useful as guidelines and help quantify the amount
of mass segregation.
Just as in the previous cases, the drop in the luminosity function at the faintest magnitudes,
especially in the central region,
could signal a problem with completeness corrections. 
The slope values derived from fits to the luminosity functions in Fig. 17 are shown
in Fig. 18. Considered individually, they are somewhat 
dependent on the adopted fit interval, but
the steepening in the luminosity functions is continuous and 
in the case of NGC 2209 extends out to the largest radius studied.

\subsection{Hodge 14}

Similarly to NGC 2209, Hodge 14 is also a low density and 
low-surface brightness cluster 
($\sigma (0) \simeq 12$ stars/pc$^{2}$; $\mu_V \simeq 21$ mag/arcsec$^{2}$).
The profiles are shown in Fig. 19; wider bins in radius ($\Delta R = 
200$ pixels = $\simeq 5$ pc) were used to reduce noise. 
Contrary to all other clusters,
there is little or no excess of stars beyond 15 pc away from the centre of 
Hodge 14.

H14's luminosity functions are noisier than the previous ones. As in NGC 2209, a power-law
in luminosities is a poor description of most luminosity function shapes seen in Fig. 20.
Yet, with the wider binning
used, it is possible to see that its luminosity function slopes display a similar behaviour
to the other clusters: a continuous rise out to some radius (in H14's case,
$R \simeq 10-12$ pc) and a drop or flattening beyond that point.
(Figs. 20 and 21).

\vskip 1. true cm

The structural parameters derived from the profiles of all 6
clusters discussed above are listed in Table~2.

\begin{table}
\caption[]{Cluster's structural parameters from density and surface brightness profiles.}
\label{tab2}
\small
\renewcommand{\tabcolsep}{1.1mm}
\begin{tabular}{lrrr}
\hline\hline
Cluster Name & $\sigma_0$ (stars/$pc^2$) & $\mu_V (0)$ (mag/arcsec$^2$) & $R_{hl}$ (pc) \\ \hline
\hline
NGC 1805 & 30$~~~~~~~~$ & 19.5$~~~~~~~~$ & 1.8 \cr
NGC 1818 & 36$~~~~~~~~$ & 17.9$~~~~~~~~$ & 2.6 \cr
NGC 1831 & 49$~~~~~~~~$ & 18.8$~~~~~~~~$ & 3.6 \cr
NGC 1868 & 54$~~~~~~~~$ & 19.0$~~~~~~~~$ & 1.8  \cr
NGC 2209 & 11$~~~~~~~~$ & 21.6$~~~~~~~~$ & 7.6 \cr
Hodge 14 & 12$~~~~~~~~$ & 21.1$~~~~~~~~$ & 2.9 \cr
\hline\hline
\end{tabular}
\end{table}

\section{Discussion }

Our first concern has been to check for any systematic effects in our
derived profiles and luminosity functions. There is plenty of photometric work on several of 
our clusters, but mostly from ground based observations and using
different photometric systems (Vallenari \etal 1993, Ferraro \etal 1995, 
Geisler \etal 1997).
In Fig. 22 we compare the surface brightness
profiles obtained by Elson (1991) for four of our clusters with the ones shown
in the previous section. Both sets of profiles display a general
agreement. Our NGC 1868 profile seems to systematically include more light 
beyond $R \sim 5$ pc. There are several possible reasons for this small
discrepancy such as overestimated sky subtraction in the integral photometry
by Elson (1991) or an overcorrection for completeness effects on our data.
But given the {\it entirely different photometric methods which led to 
the profiles shown in the figure, the level of agreement is evidence that
there are no strong systematic effects in the photometry, sample
selection and background subtraction in either work.}

There is a relative paucity of deep enough luminosity functions published in the literature for
LMC clusters to enable a direct comparison to our data. 
Vallenari \etal (1992) have 
obtained BV photometry for NGC 1831 down to $V$ \ltsima $22$ and derived a 
cumulative luminosity function down to $V \simeq 21$. This limit 
is close to the bright limit of our sample and the overlap in magnitudes is too
small to allow a comparison. A similar difficulty occurs with the luminosity functions for NGC 
1868 by Flower \etal (1980) and for NGC 1818 by Will \etal (1995). Olsen \etal
(1998) analyzed deep HST/WFPC2 CMDs for 6 
old globular clusters in the LMC, deriving
their ages, metallicities and horizontal branch structures, but 
did not investigate their luminosity functions. 
Elson \etal (1998c) also show an HST CMD of NGC 1818
and discuss its binary sequence but do not use the data for deriving a luminosity function.

For NGC 1818, however, Hunter \etal (1997) have obtained deep HST 
photometry, also using the F555W and F814W filters. 
Fig. 23 shows the comparison between their
global mass function and ours. The conversion from luminosity to mass
in our sample was done using a suitable isochrone from Worthey \etal (1999)
in a similar way as in Santiago \etal (1996) and Santiago \etal (1998).
We do not cover the high mass end of the Hunter \etal data since we are again
restricting the mass range to that sampled by the deep (large exposure time)
HALF field.
But the agreement between the two mass functions in the common mass range is
striking. We thus, conclude that the present data, besides
making up a homogeneous dataset, lack any strong systematics due to sample
selection or completeness.

How similar are the mass functions and their systematic trend with position 
among our sample clusters? Are our data consistent with a universal shape
for the IMF? These questions are not easily answered without careful
attention to two important issues: a) Derivation of present day mass functions
(PDMFs) from luminosity functions require a mass-luminosity relation, which in turn depends
on metallicity and age. Thus, this question is intimately related to finding
the best fitting isochrones to the cluster CMDs 
and checking their sensitivity to different 
evolutionary models. This is currently underway for our clusters CMDs;
b) Recovering the IMF from the PDMF 
requires reconstruction of the dynamical history of the
cluster, something more easily done with the help of N-body simulations.

At present, however, we concentrate on the 2 youngest clusters in the
sample, NGC 1805 and NGC 1818, whose ages are below $\tau \simeq
40\times10^6$ years. Fig. 24 shows the global mass functions for both clusters
(panel a), plus the mass functions in an outer region (panel b). The
two mass functions differ, that of NGC 1805 being steeper. These mass
functions likely reflect the IMF in both cases, especially for the
outer radius shown, where the dynamical timescales for $M$
\ltsima $3~M_{\odot}$ stars are \gtsima $10^9$ yrs, somewhat longer
than the clusters ages. For the outer region, the mass function slopes, 
$\Gamma = d log \phi (M) / d log (M)$, are $-1.5$ for NGC 1818 and
$-2.4$ for NGC 1805, therefore steeper than the Salpeter value of
$\Gamma = -1.35$.  This result suggests that the shape of the the IMF
may vary with environment or metallicity (see Scalo (1998) for a
recent review on IMF variations).

Note however, that Elson \etal (1998c) have analyzed the
binary fraction of NGC 1818 using HST V \vs U-V data
and have found it to increase inwards. The authors argue that this is
consistent with dynamical evolution.
In fact, unresolved binaries may have an effect on the observed luminosity function or mass function
shape. We are currently using statistical tools and modelling techniques
that make use of the full two-dimensional information available from the 
cluster CMDs, in order to derive binary fractions and mass function shapes simultaneously
for each cluster, and investigate their variation with distance from the 
centre.

\section{Summary}

We have used extensive WFPC2 photometry of stars
in 6 rich star clusters in the LMC
in order to obtain and analyze their density and surface brightness
profiles, derive structural parameters and
investigate the dependence of their luminosity functions with position.
The cluster sample was chosen to display large variations in age, metallicity
and structural parameters (Elson \etal 1998a,b, Beaulieu \etal 2000).

We obtain half-light radii and central stellar densities
directly from the clusters density and surface brightness profiles. 
The clusters studied here display variable profile shapes, sizes and central
densities. Our star number surface density and surface brightness
profiles, despite being constructed by adding the individual contribution
of each star within the cluster sample, show a general agreement with
profiles obtained with integral ground-based photometry.

The luminosity functions (within the range $0.5$ \ltsima $M_V$ \ltsima $5.5$, $3$ \gtsima
$M/M_{\odot}$ \gtsima $0.9$) of 
our clusters display a systematic dependence with distance from 
the cluster centre, in the sense of being steeper at larger radii. This mass
segregation is seen even in the two youngest clusters in our sample: NGC 1805
and NGC 1818. The ages of these two clusters are $\tau$ \ltsima $40$ Myrs.
Assuming a central velocity dispersion in these clusters of $\sigma_0 
\simeq 1$ \kms, the two-body relaxation time for $3~M_{\odot}$
stars at $r \simeq 2$ pc from the centre (mid-distance of 
the most central luminosity function ring used) would be
\gtsima 100 Myrs. Thus, at least for these clusters, these trends 
in luminosity function shape may be primordial, something which would have
important implications for the physics of star formation within protocluster
gas clouds. We converted their luminosity functions into mass functions and made a direct comparison 
between the two: we find that NGC 1805 has a systematically steeper mass 
function than NGC 1818, both globally and in the outer regions, where 
dynamical timescales are likely to be longer than 1 Gyr.

Systematic steepening in the luminosity function with radial distance is also observed in 
NGC 1831, NGC 1868, NGC 2209 and H14. Luminosity function shapes vary considerably from
cluster to cluster or even from one region to another in the same cluster.
The luminosity functions are in many cases not well described by a straight line
in the $log \Phi \vs M_{555}$ plane (a power-law in luminosity). 
However, slopes derived from linear fits still usefully
trace the more significant changes in luminosity function shape and in the
relative proportions of high and low mass main-sequence stars.

Steepening in the luminosity function from the cluster centre outwards is expected due to 
mass segregation caused by energy equipartition. N-body modelling of cluster
sized samples of point masses also reproduce this trend. It would be
very interesting to directly compare model predictions with the observed
positional dependent luminosity functions (or mass functions) in this work. Vesperini \& Heggie
(1997) have modelled the evolution of the cluster stellar mass function, taking
into account relaxation, stellar evolution and disk shocking and tidal effects
due to the Galaxy. Their simulations reconstruct the evolution of the mass 
function slope for different parts of a cluster. Unfortunately, the mass
range probed in their simulations does not coincide with ours, preventing
a straight comparison between the predicted slope variations and the observed
ones.

We compared our global mass function obtained for NGC 1818 
with that of Hunter \etal 
(1997) and found excellent agreement between the two. Unfortunately, to our
knowledge, this is the only deep enough mass function determination from other
authors to allow a direct comparison.

We are currently analyzing the colour magnitude diagrams of our clusters
using objective statistical tools in an attempt to extract from the data
the most reliable values for cluster age and metallicity. These in turn 
will allow more reliable conversion of the clusters luminosity functions into mass functions, making
it easier to compare with the results of dynamical modelling. Mass functions
and binary fractions are also currently being obtained from these data,
by statistically comparing observed CMDs and luminosity functions to model ones.

\begin{acknowledgements}
We are deeply indebted to the late Rebecca Elson, not only for the crucial role
she played in planning and carrying out this project, 
but also for the great deal we learned from her.
BXS wishes to thank the hospitality and support of the IoA, in Cambridge, UK.
\end{acknowledgements}

%
%

\vfill\eject
$~~~~~~~~~~$
\vfill\eject

\noindent
{\bf Figure Captions} 

\medskip
\noindent
{\bf Fig. 1.} 
Photometric consistency check for NGC 1805. $m_{814}$ and $m_{555}$ magnitudes
measured in both CEN and HALF fields are compared on the upper panel.
The two close straight lines in each case represent the identity function 
and a linear regression fit. The lower panel shows the comparison between
CEN and HALF $m_{555} - m_{814}$ colours, again with the identity line for
guidance.

\medskip
\noindent
{\bf Fig. 2.} 
Completeness values for NGC 1868 CEN field, chip 1 (PC), $23 < m_{814} < 
23.5$, calculated as described in Sect. 2.3. The completeness
map is shown as percentage values superposed to the distribution of
automatically detected objects on the chip. 

\medskip
\noindent 
{\bf Fig. 3.} 
Distribution on the plane of the sky of stars in NGC 1805. The crosses
(triangles) represent stars in the CEN (HALF) field, the cluster centre
coinciding with the CEN field PC chip.

\medskip
\noindent
{\bf Fig. 4.} 
Stellar surface density (panel {\it a}) and surface brightness
(panel {\it b}) profiles for NGC 1805. The upper curves are not corrected
for background star contamination. The corrected profiles are the lower curves.

\medskip
\noindent
{\bf Fig. 5.} 
Positional dependent luminosity functions for NGC 1805. The range in radius
to which each luminosity function corresponds is indicated in each panel (1~pc = 41 pix). 
The luminosity functions have been corrected for background contamination as explained in the
text. Linear fits to the luminosity functions are shown on each panel. The best fit lines
bracket the range in $M_{555}$ used in the fit.

\medskip
\noindent
{\bf Fig. 6.} 
Slope values obtained from linear fits to the luminosity functions shown in 
Fig. 5 as a function of distance from cluster's centre.

\medskip
\noindent
{\bf Fig. 7.} 
Stellar surface density (panel {\it a}) and surface brightness
(panel {\it b}) profiles for NGC 1818. The upper curves are not corrected
for background star contamination. The corrected profiles are the lower curves.

\medskip
\noindent
{\bf Fig. 8.} 
Positional dependent luminosity functions for NGC 1818. The range in radius
to which each luminosity function corresponds is indicated in each panel (1~pc = 41 pix). 
The luminosity functions have been corrected for background contamination as explained in the
text. Linear fits to the luminosity functions are shown on each panel. The best fit lines
bracket the range in $M_{555}$ used in the fit.

\medskip
\noindent
{\bf Fig. 9.} 
Slope values obtained from linear fits to the luminosity functions shown in 
Fig. 8 as a  function of distance from cluster's centre.

\medskip
\noindent
{\bf Fig. 10.} 
Stellar surface density (panel {\it a}) and surface brightness
(panel {\it b}) profiles for NGC 1831. The upper curves are not corrected
for background star contamination. The corrected profiles are the lower curves.

\medskip
\noindent
{\bf Fig. 11.} 
Positional dependent luminosity functions for NGC 1831. The range in radius
to which each luminosity function corresponds is indicated in each panel (1~pc = 41 pix). 
The luminosity functions have been corrected for background contamination as explained in the
text. Linear fits to the luminosity functions are shown on each panel. The best fit lines
bracket the range in $M_{555}$ used in the fit.

\medskip
\noindent
{\bf Fig. 12.} 
Slope values obtained from linear fits to the luminosity functions shown in 
Fig. 11 as a  function of distance from cluster's centre.

\medskip
\noindent
{\bf Fig. 13.} 
Stellar surface density (panel {\it a}) and surface brightness
(panel {\it b}) profiles for NGC 1868. The upper curves are not corrected
for background star contamination. The corrected profiles are the lower curves.

\medskip
\noindent
{\bf Fig. 14.} 
Positional dependent luminosity functions for NGC 1868. The range in radius
to which each luminosity function corresponds is indicated in each panel (1~pc = 41 pix). 
The luminosity functions have been corrected for background contamination as explained in the
text. Linear fits to the luminosity functions are shown on each panel. The best fit lines
bracket the range in $M_{555}$ used in the fit.

\medskip
\noindent
{\bf Fig. 15.} 
Slope values obtained from linear fits to the luminosity functions shown in 
Fig. 14 as a  function of distance from cluster's centre.

\medskip
\noindent
{\bf Fig. 16.} 
Stellar surface density (panel {\it a}) and surface brightness
(panel {\it b}) profiles for NGC 2209. The upper curves are not corrected
for background star contamination. The corrected profiles are the lower curves.

\medskip
\noindent
{\bf Fig. 17.} 
Positional dependent luminosity functions for NGC 2209. The range in radius
to which each luminosity function corresponds is indicated in each panel (1~pc = 41 pix). 
The luminosity functions have been corrected for background contamination as explained in the
text. Linear fits to the luminosity functions are shown on each panel. The best fit lines
bracket the range in $M_{555}$ used in the fit.

\medskip
\noindent
{\bf Fig. 18.} 
Slope values obtained from linear fits to the luminosity functions shown in 
Fig. 17 as a function of distance from cluster's centre.

\medskip
\noindent
{\bf Fig. 19.} 
Stellar surface density (panel {\it a}) and surface brightness
(panel {\it b}) profiles for Hodge 14. The upper curves are not corrected
for background star contamination. The corrected profiles are the lower curves.

\medskip
\noindent
{\bf Fig. 20.} 
Positional dependent luminosity functions for Hodge 14. The range in radius
to which each luminosity function corresponds is indicated in each panel (1~pc = 41 pix). 
The luminosity functions have been corrected for background contamination as explained in the
text. Linear fits to the luminosity functions are shown on each panel. The best fit lines
bracket the range in $M_{555}$ used in the fit.

\medskip
\noindent
{\bf Fig. 21.} 
Slope values obtained from linear fits to the luminosity functions shown in 
Fig. 20 as a function of distance from cluster's centre.

%
%
%

\medskip
\noindent
{\bf Fig. 22.} 
Background corrected surface brightness profiles 
for 4 clusters in our sample compared to those
from Elson (1991), based on integral photometry. The open circles
are the data from Elson (1991), while the solid points are from the present
paper.

\medskip
\noindent
{\bf Fig. 23.} 
Comparison of the NGC 1818 mass function 
from this work to that of Hunter \etal (1997).
The symbols are as indicated.

\medskip
\noindent
{\bf Fig. 24.} 
Comparison between the NGC 1805 and NGC 1818 mass functions. Panel {\it a)}
The global mass functions, within the range $0 \leq R \leq 25$ pc. Panel {\it b)} The
mass functions for an outer radius, as indicated.

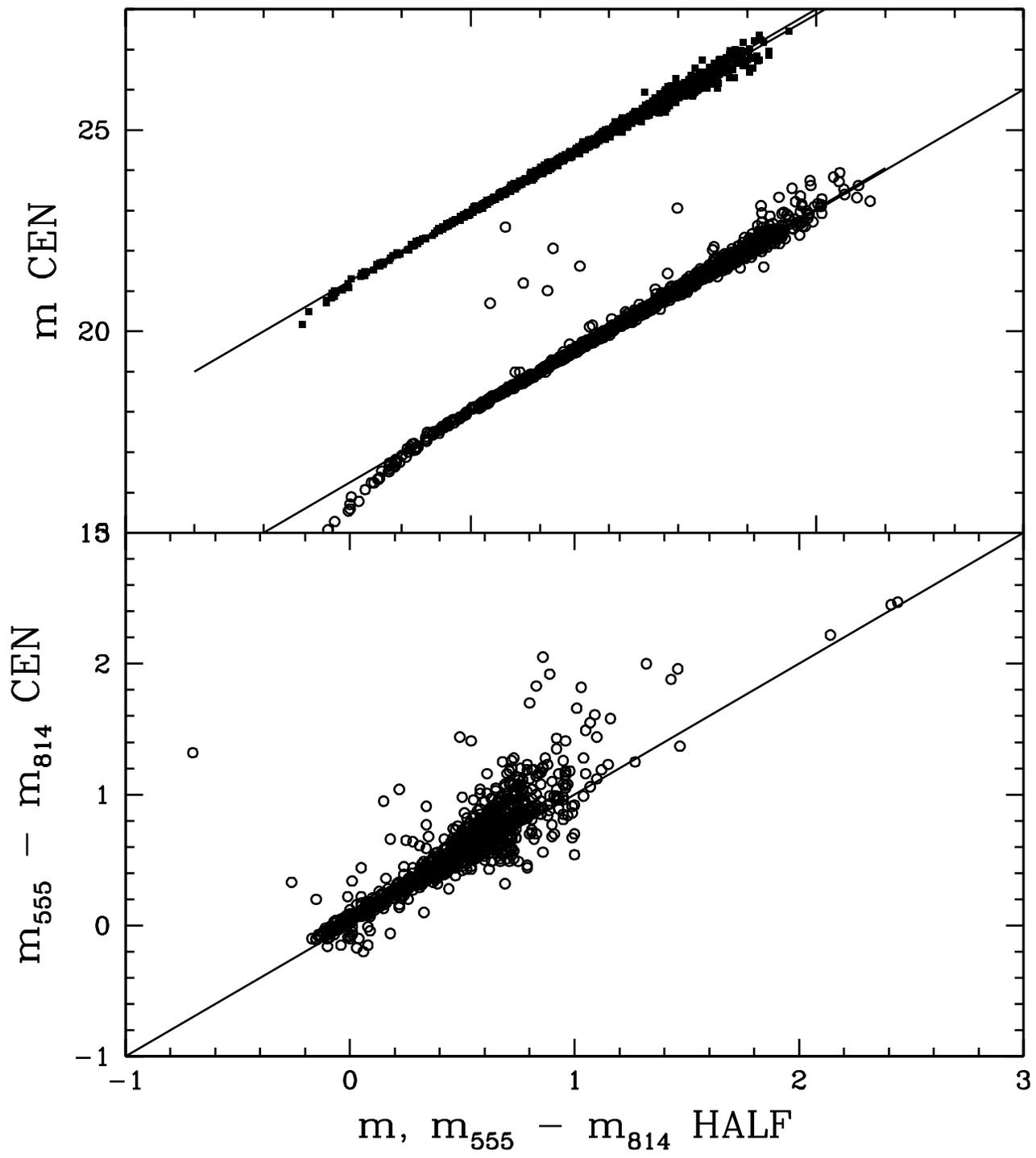
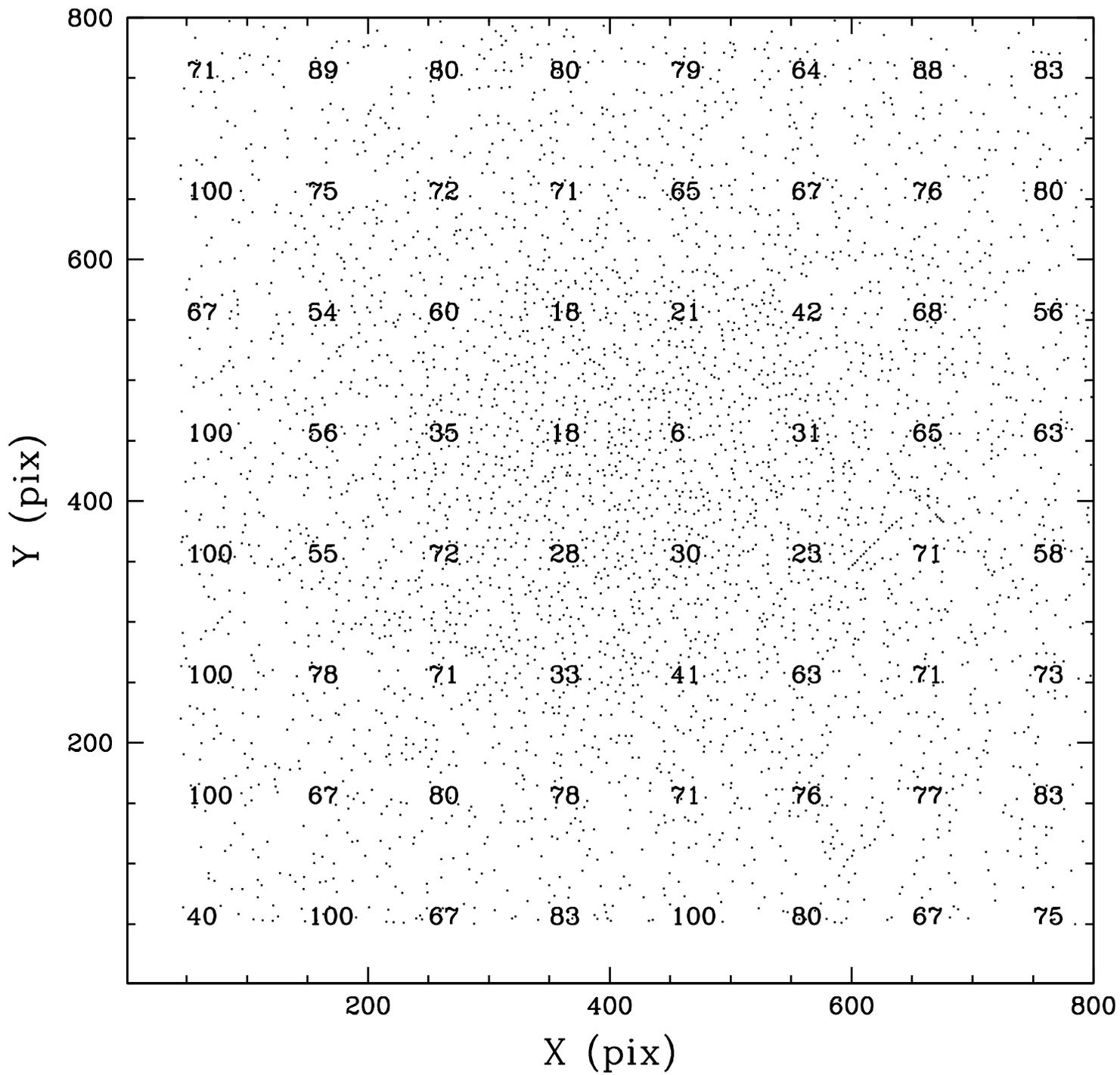
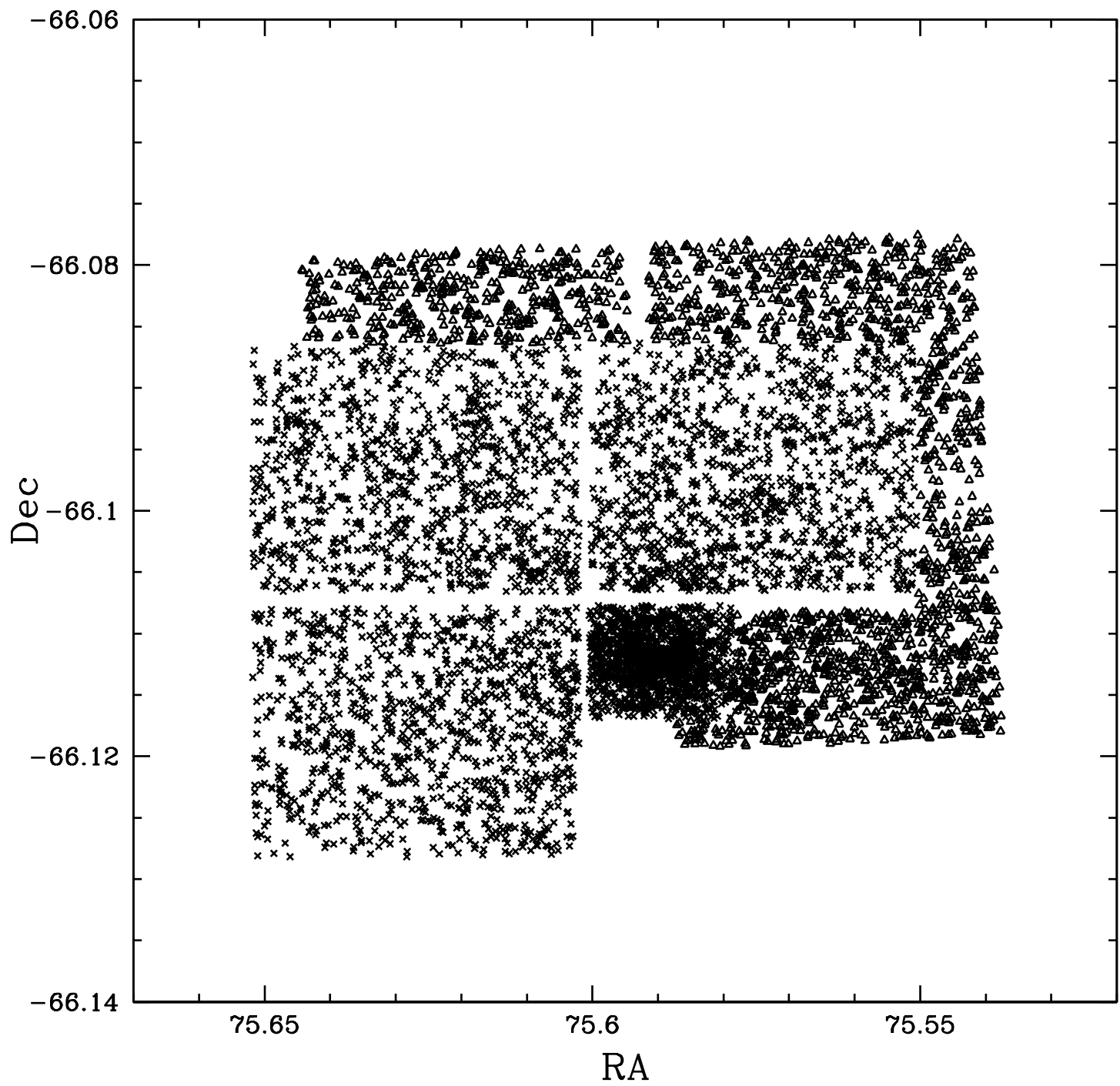
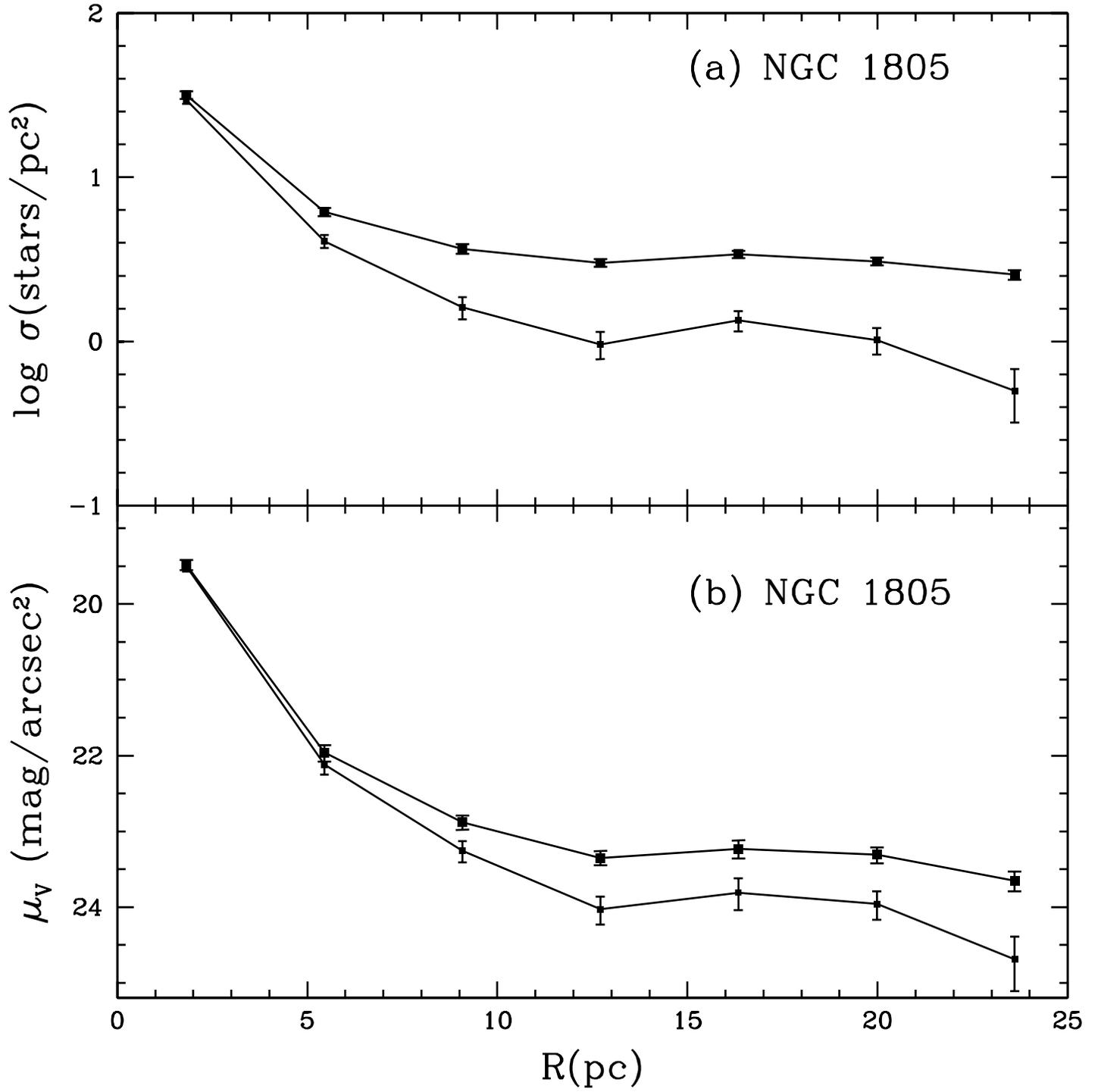
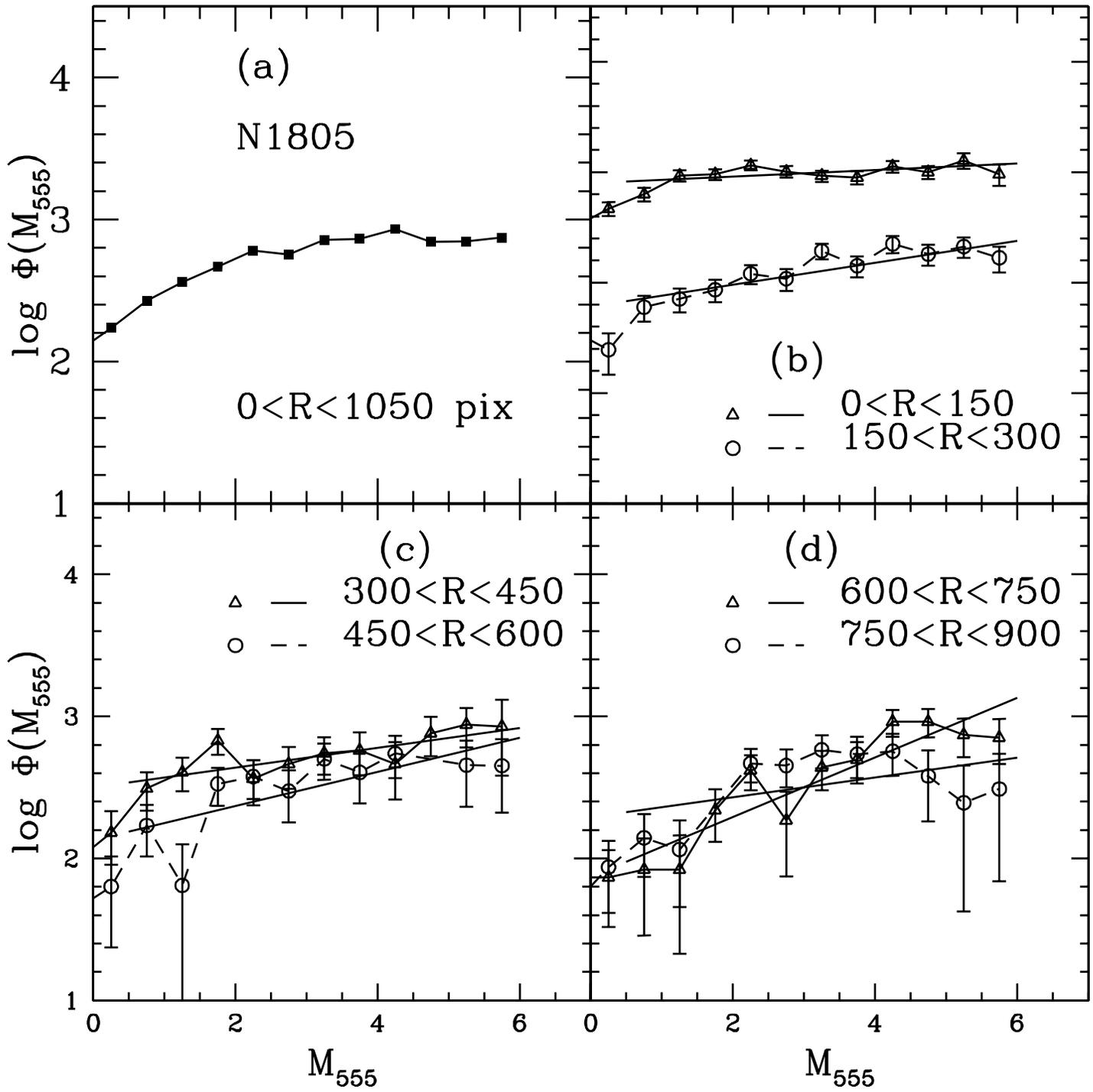
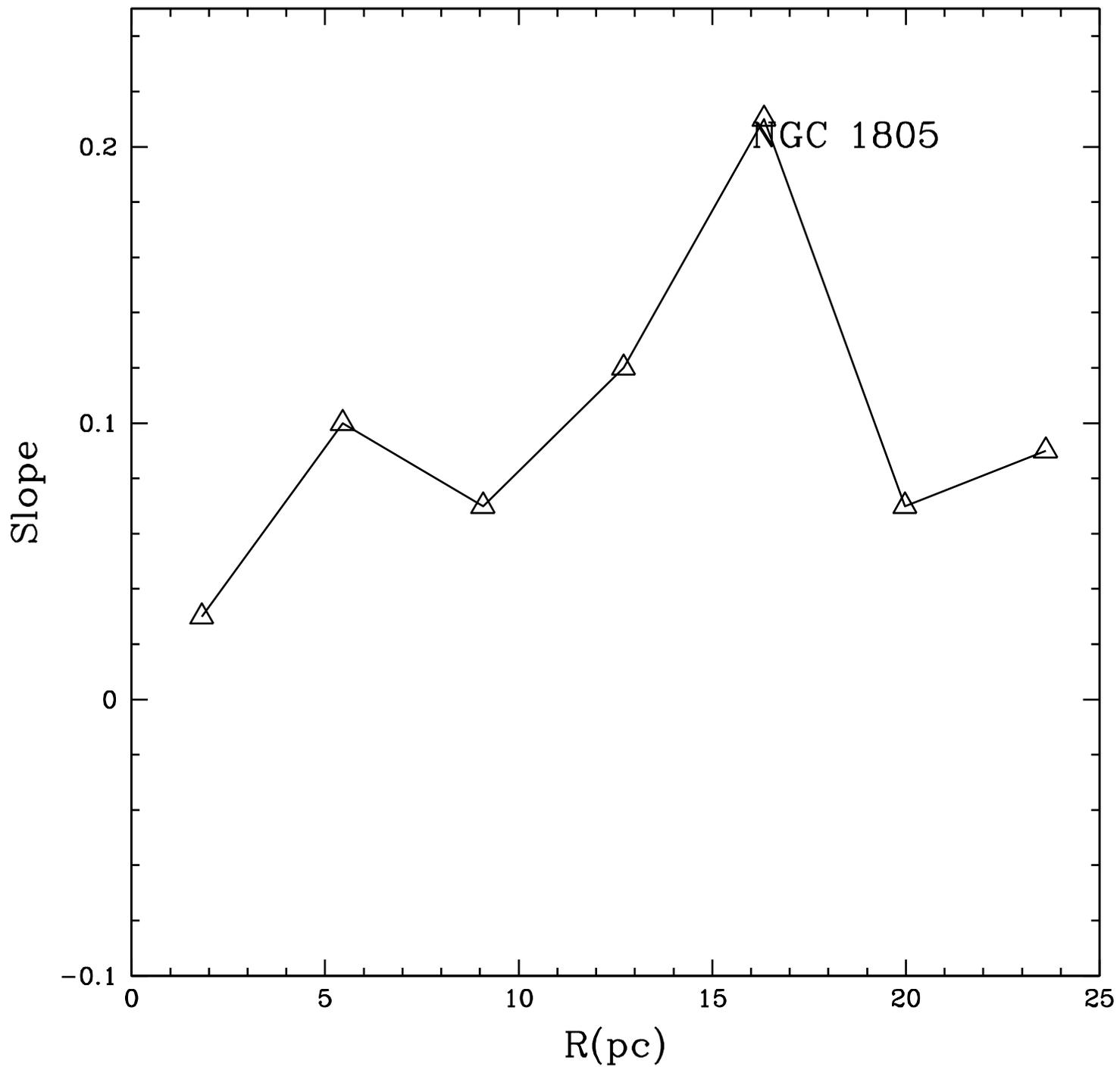
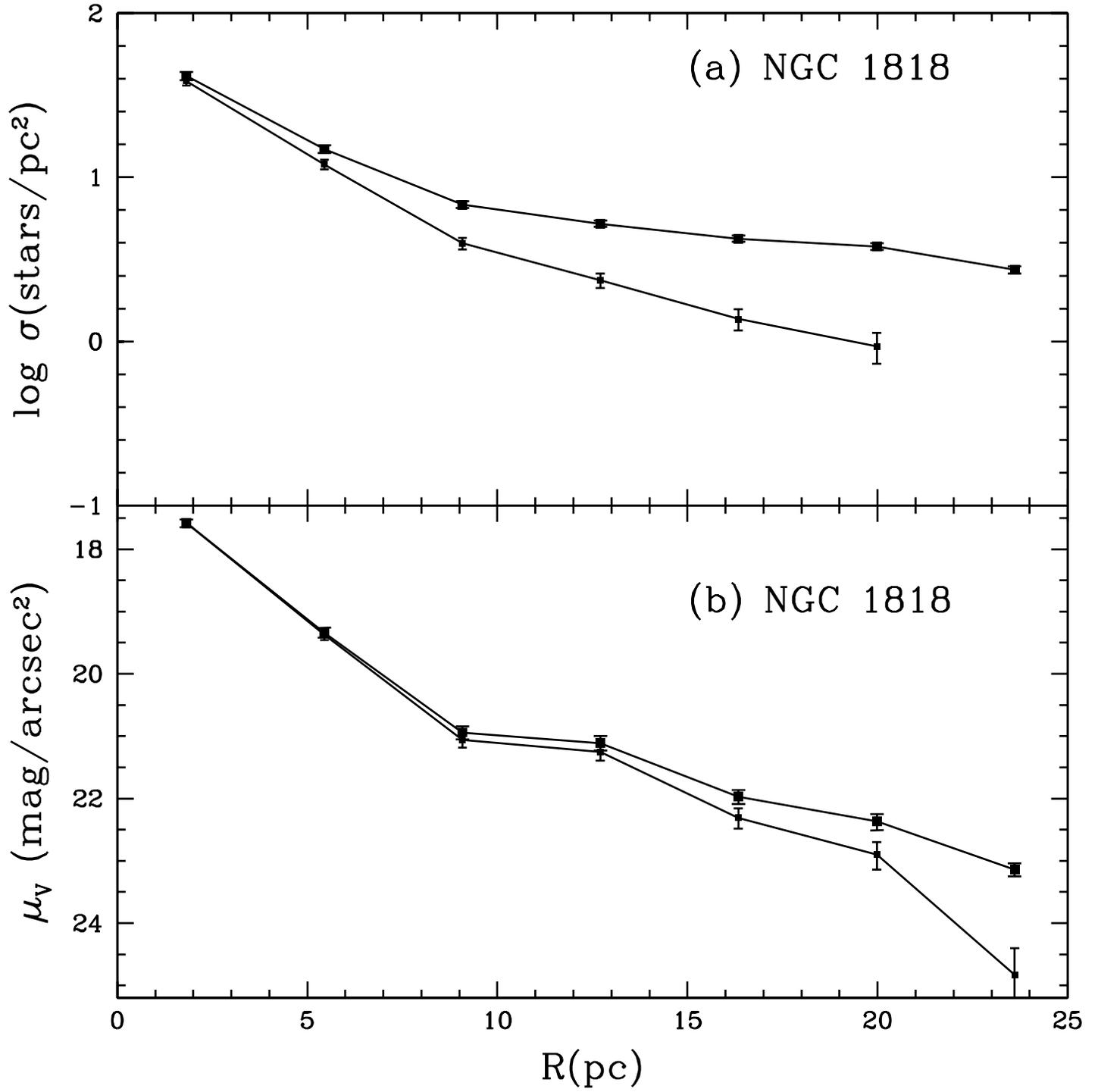
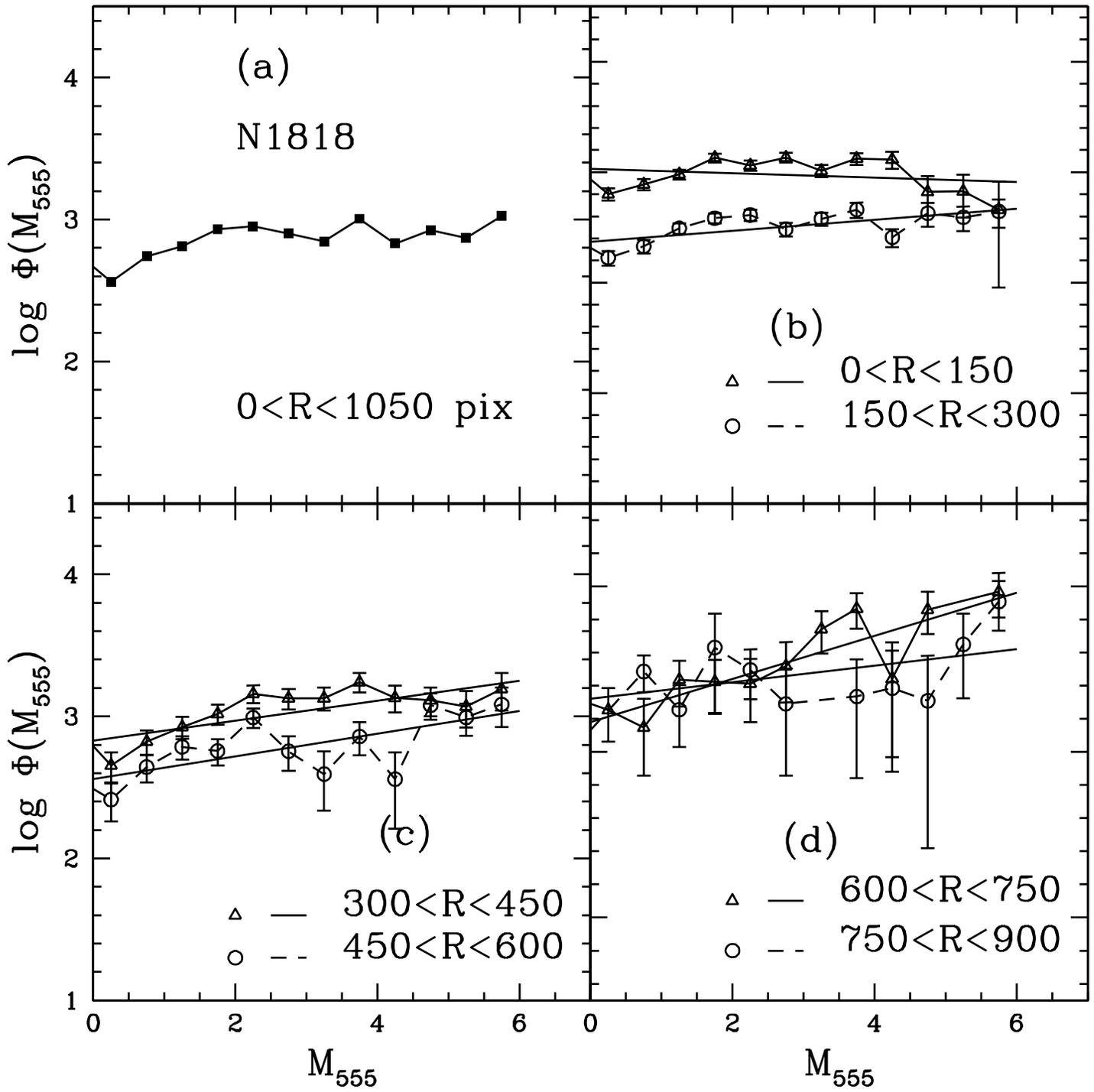
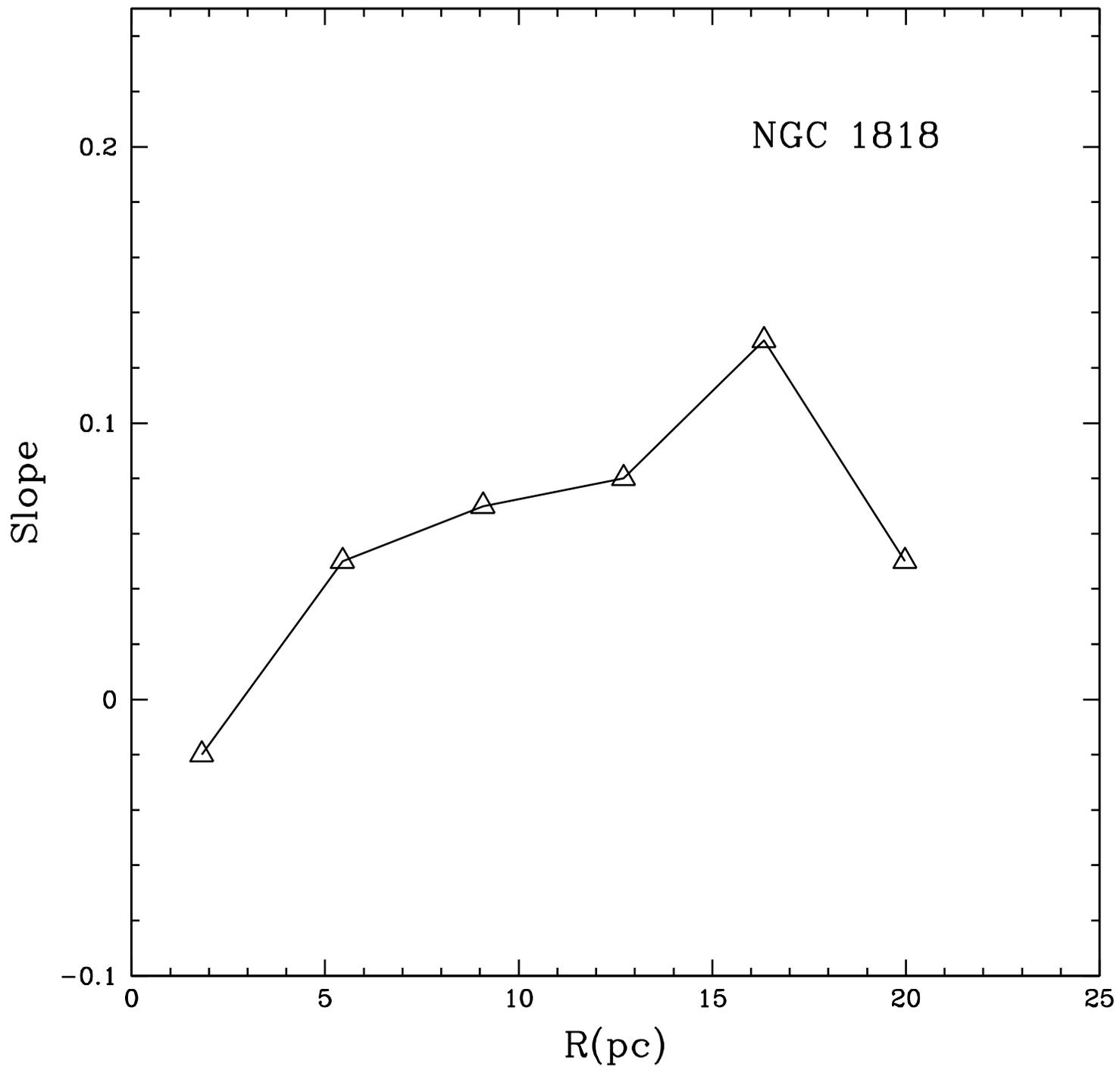
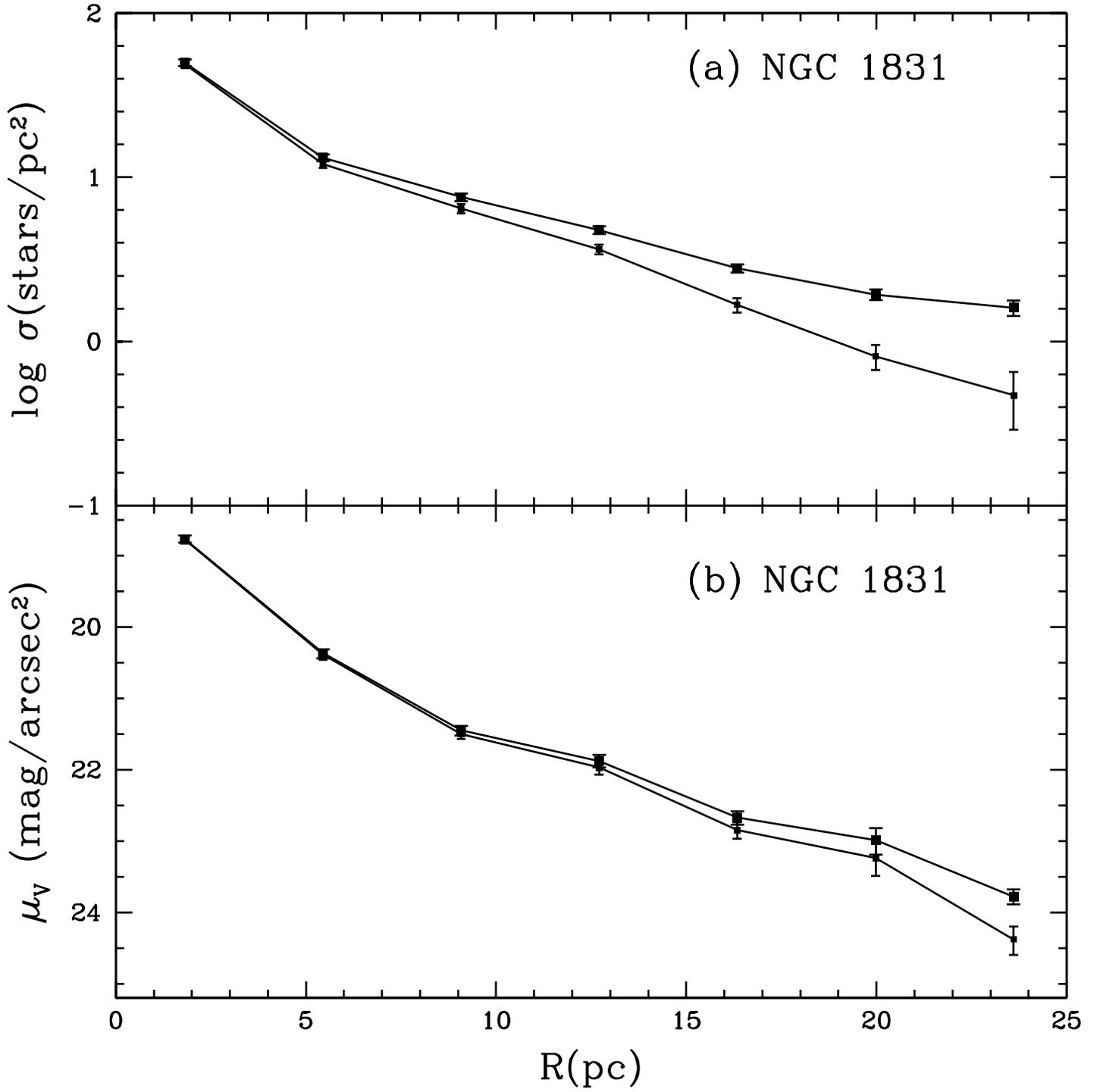
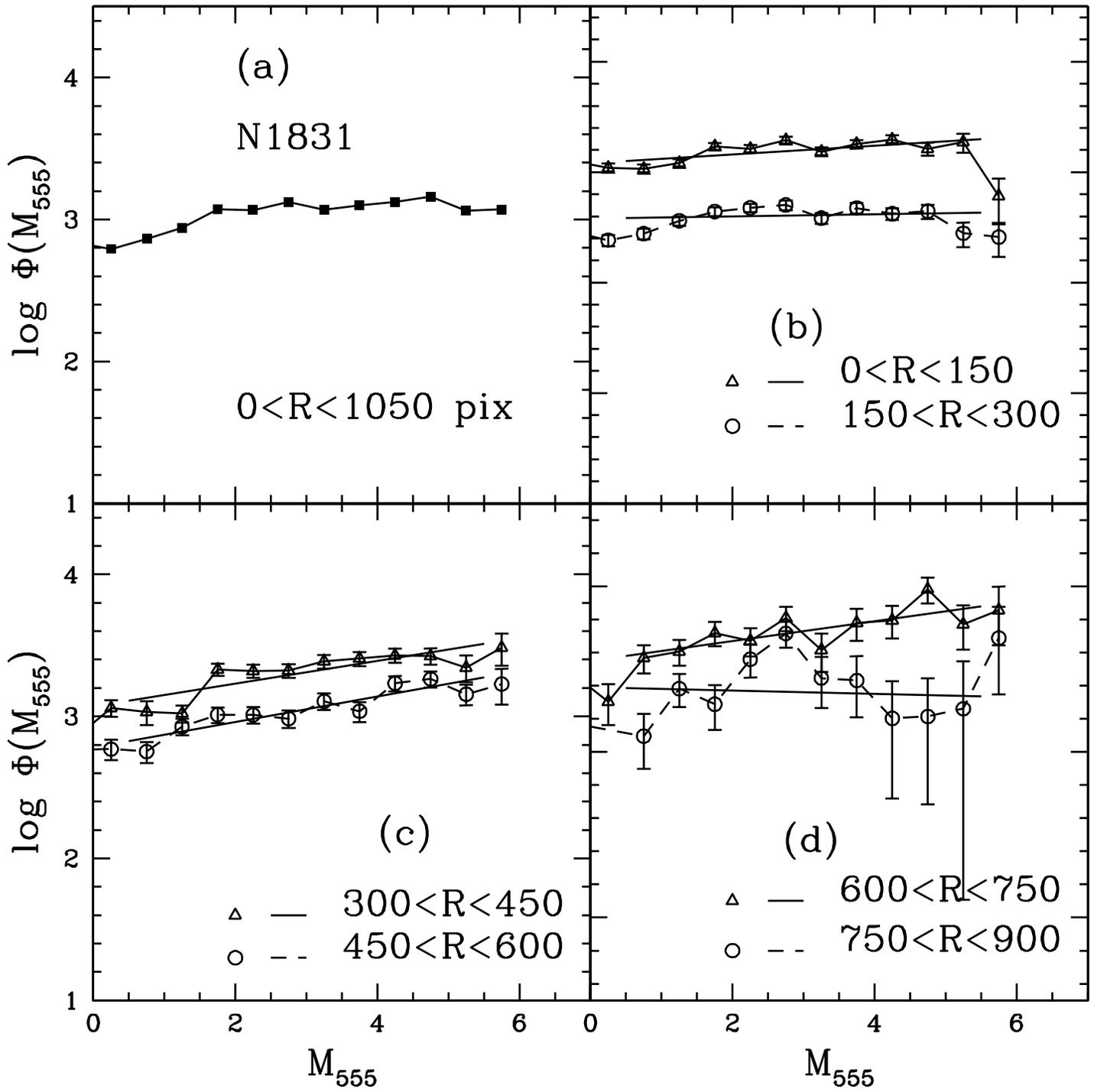
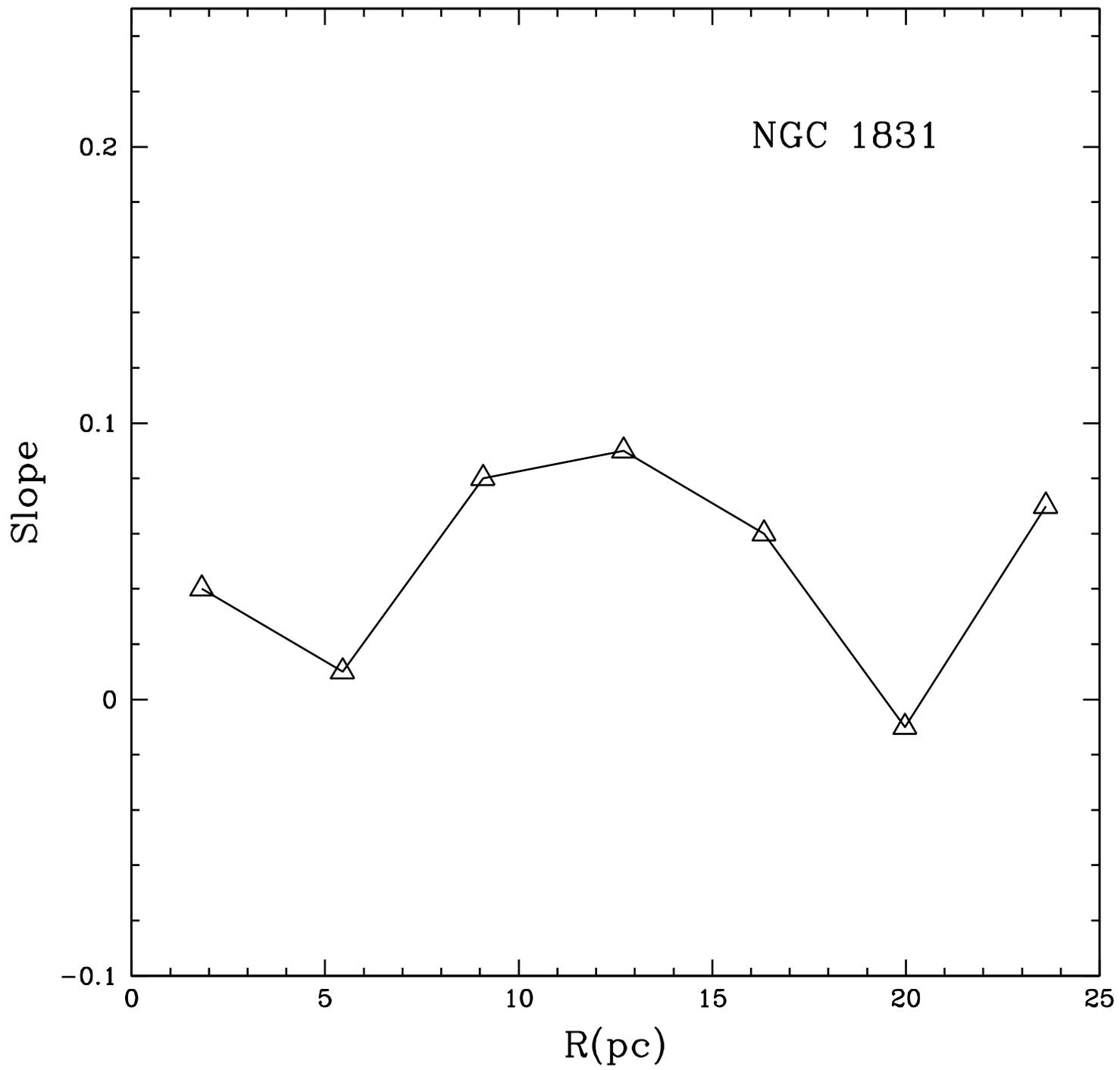
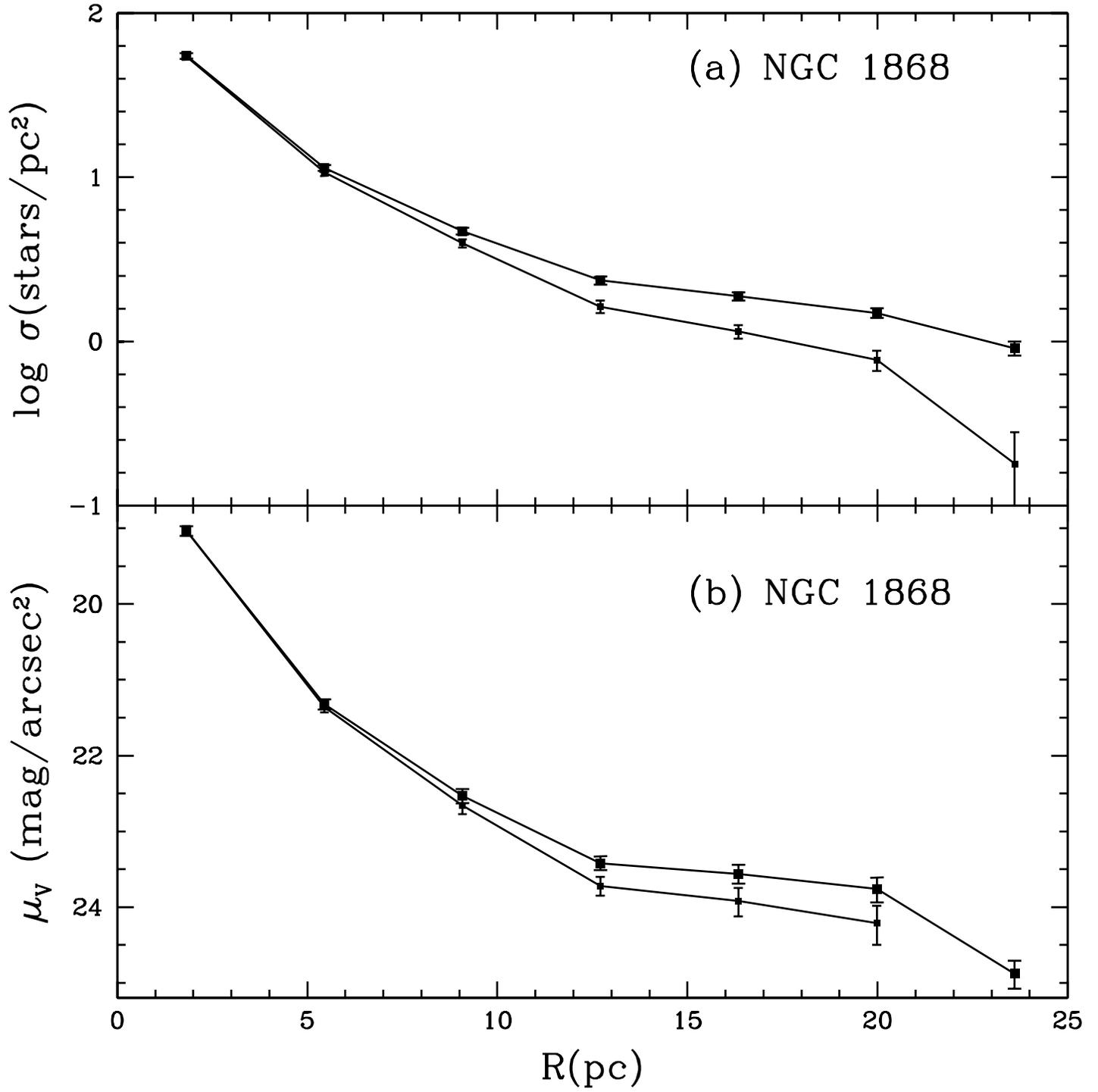
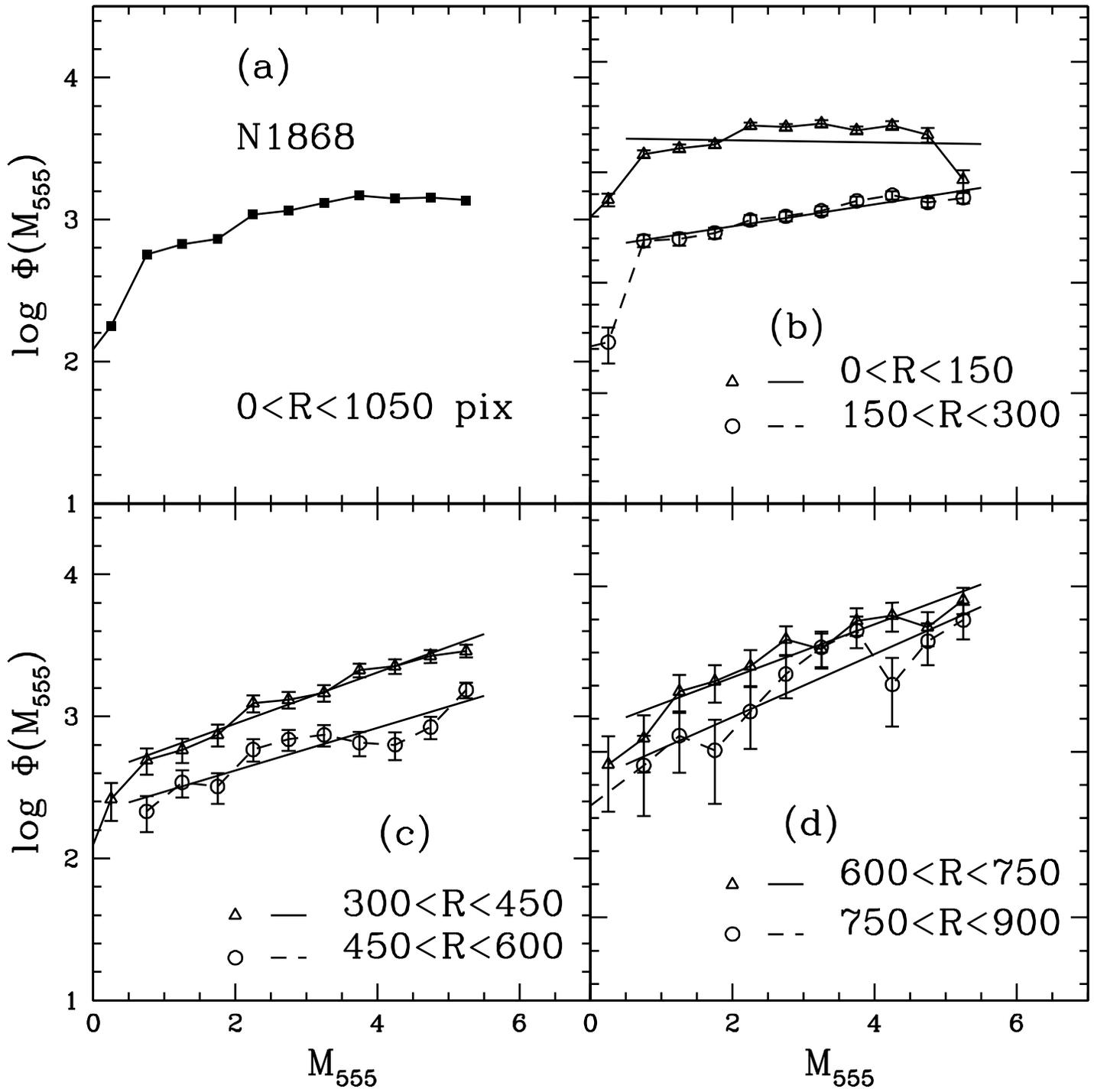
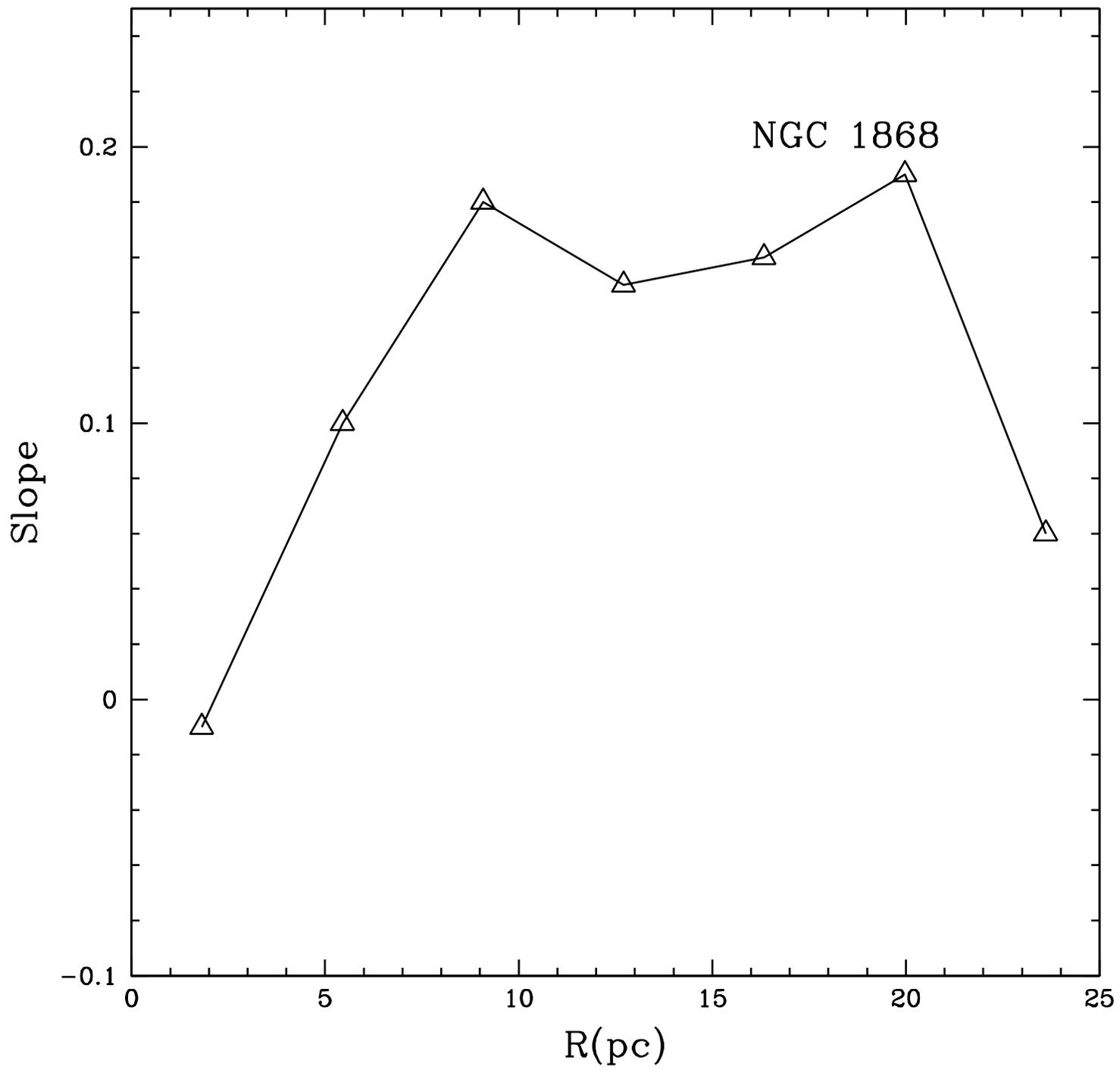
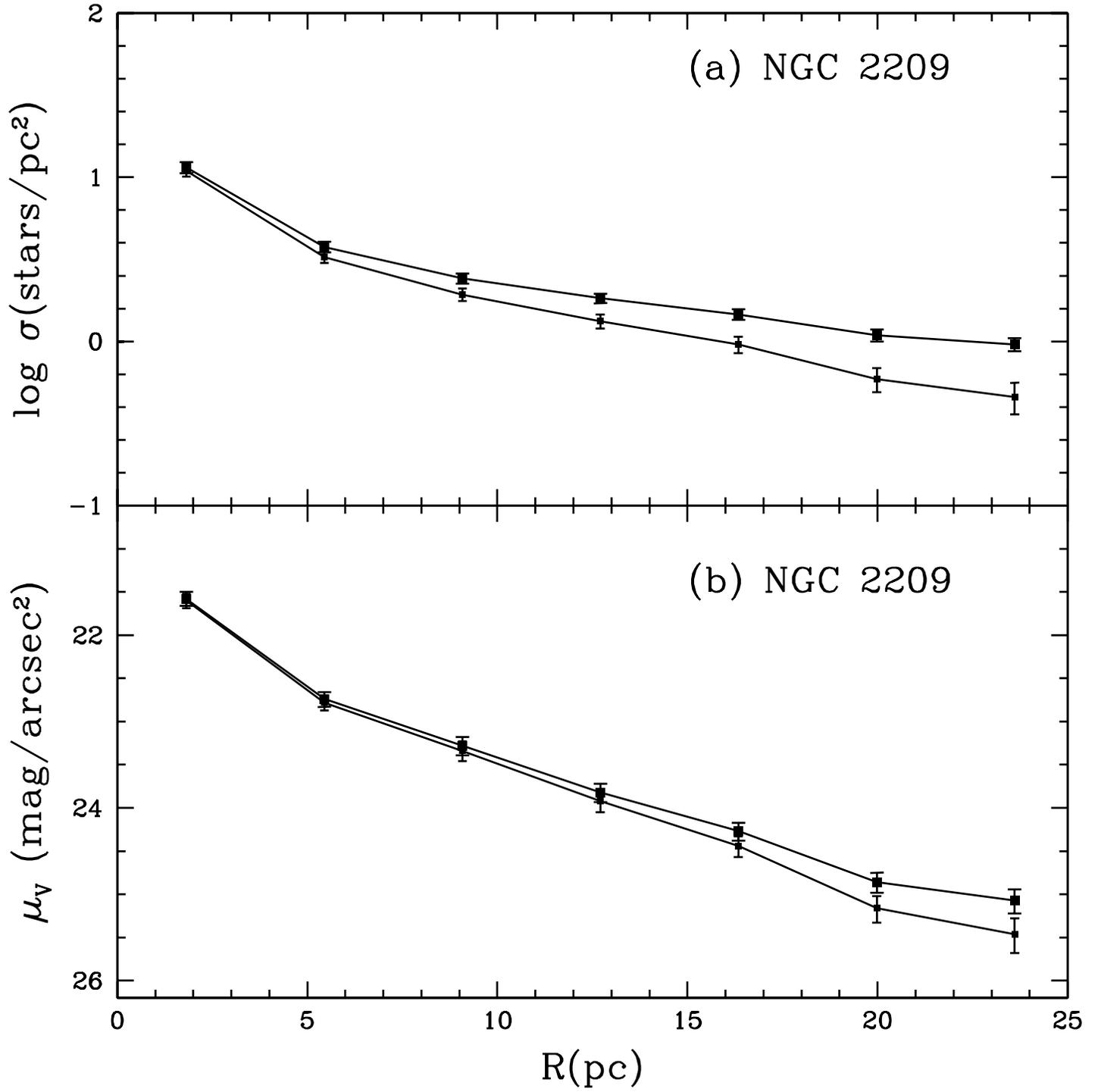
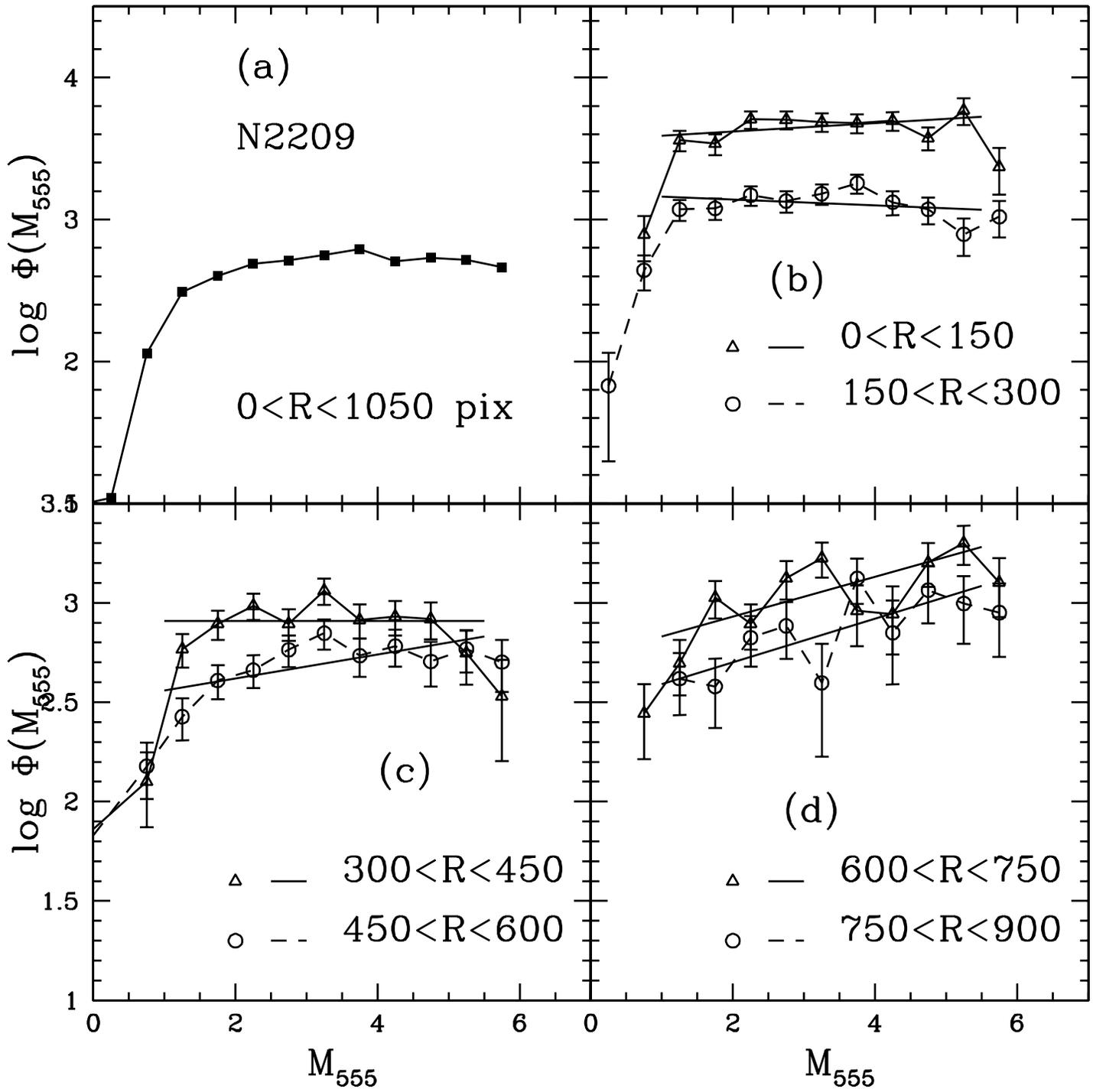
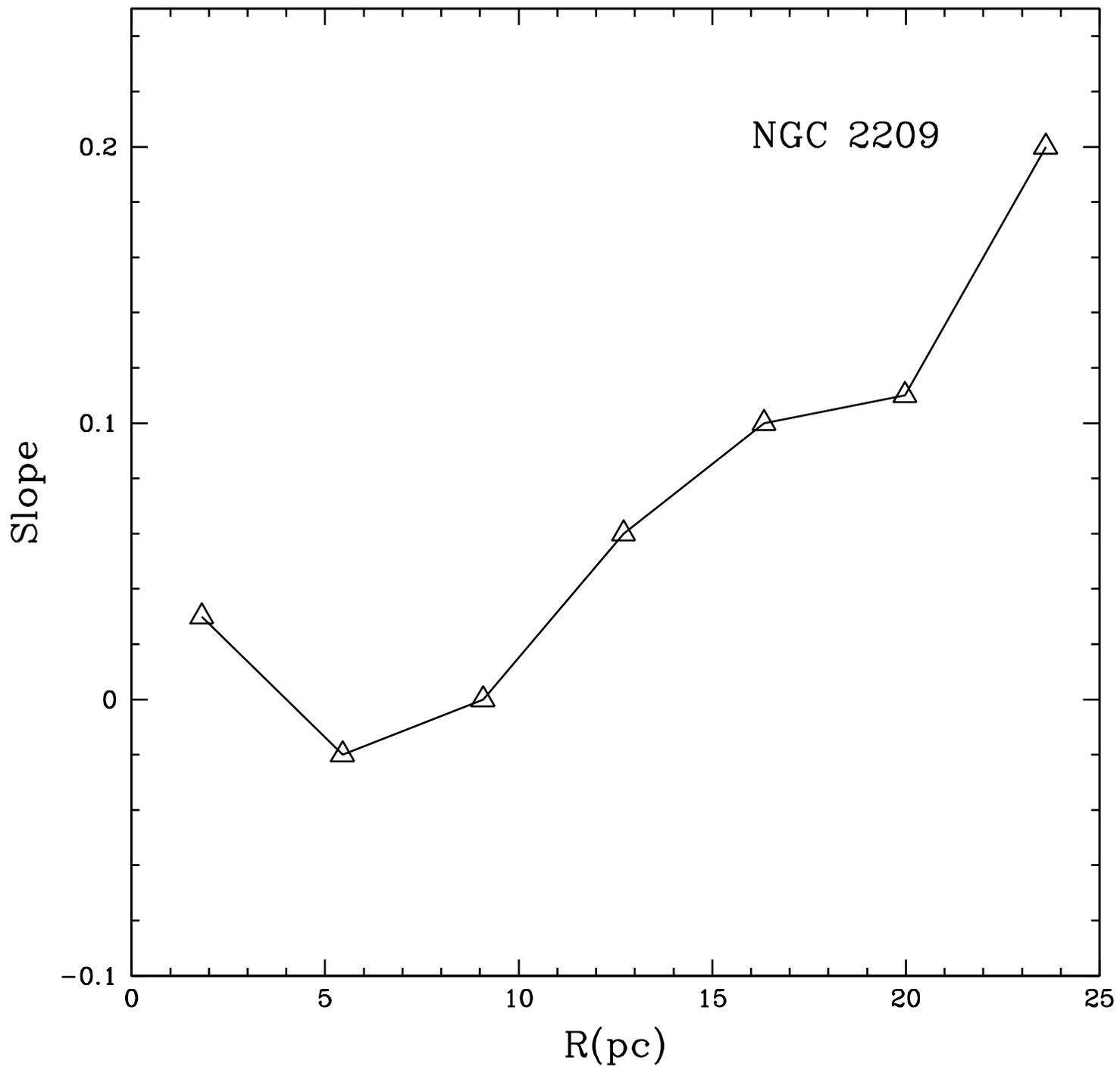
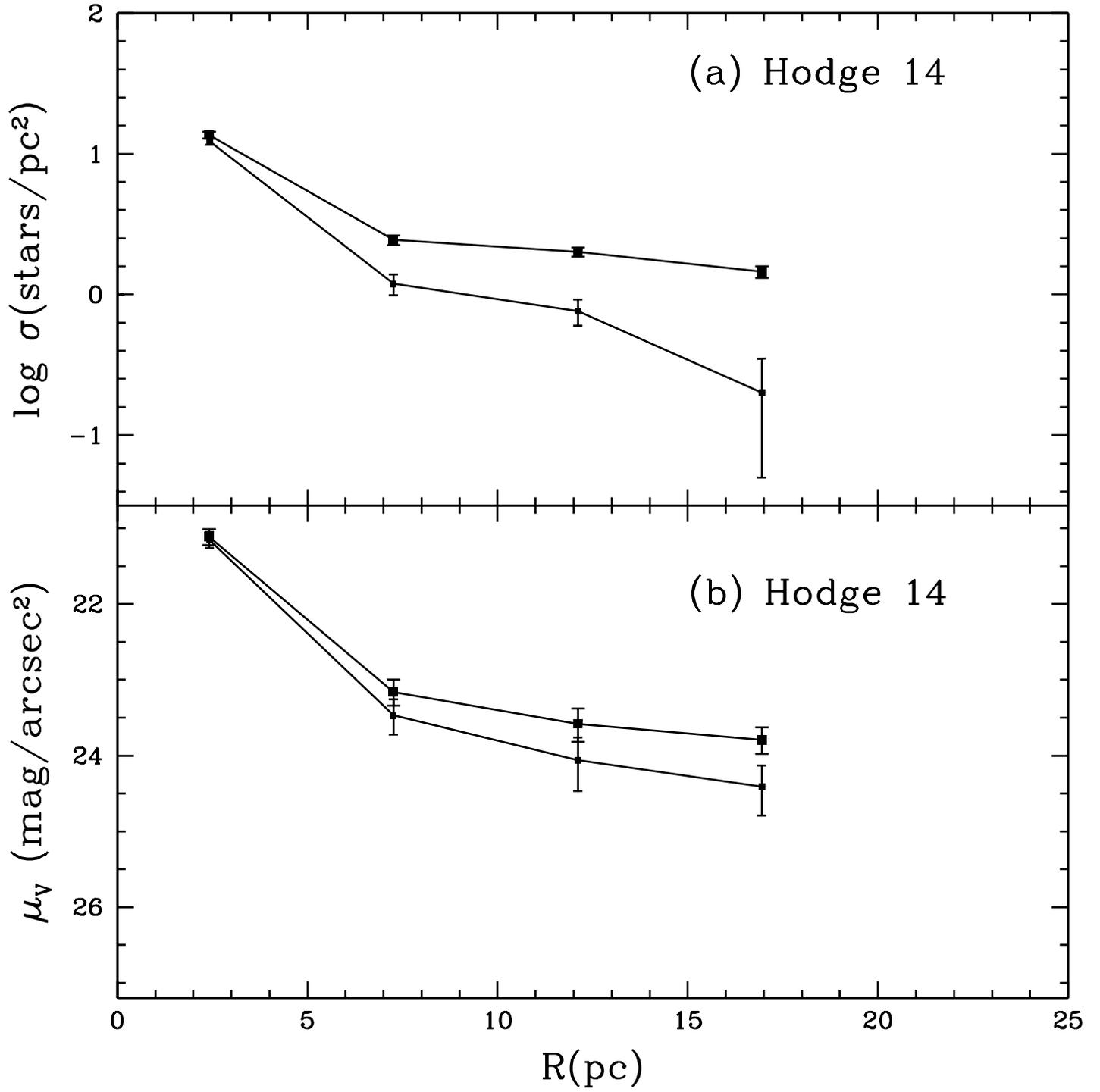
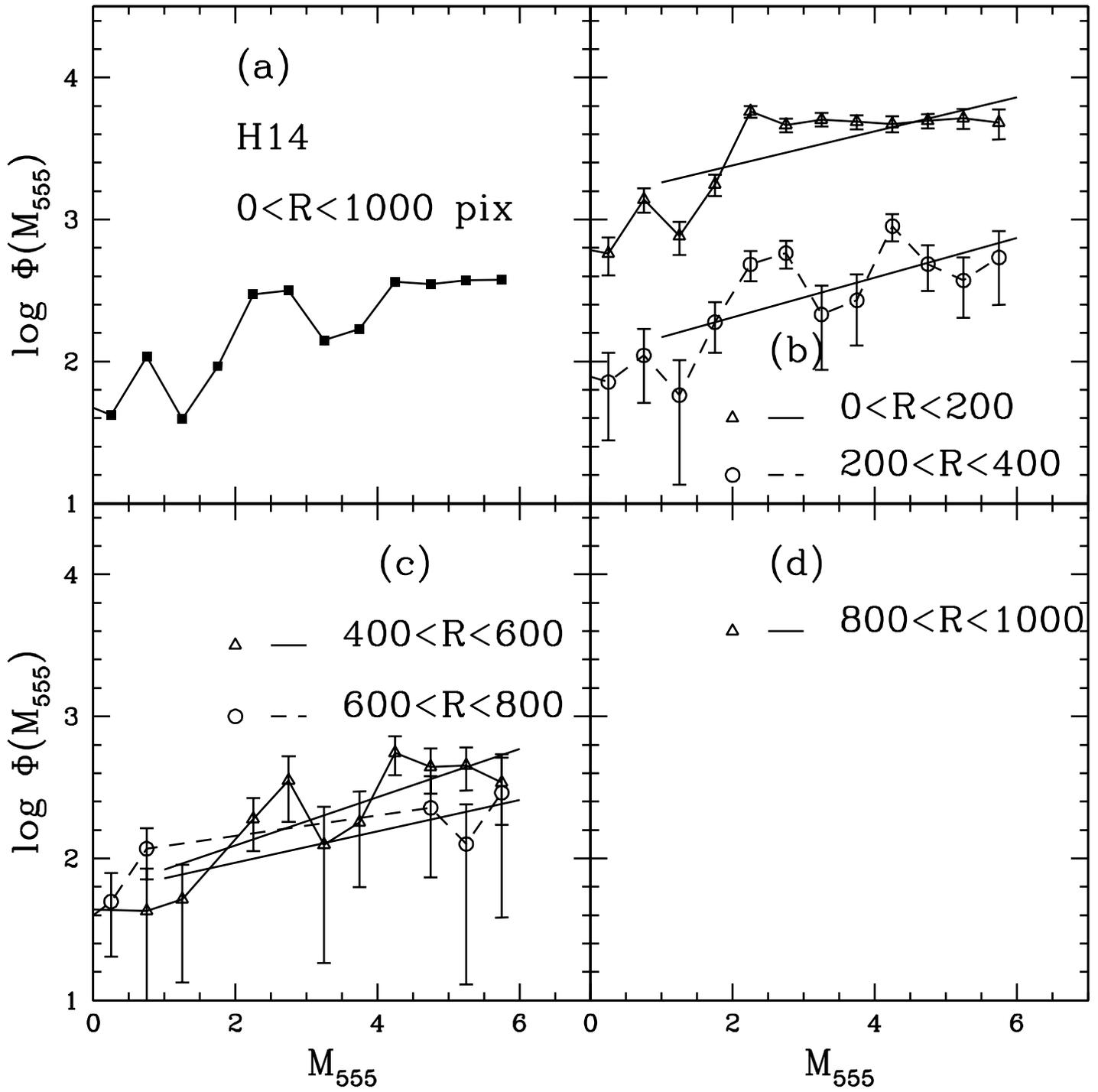
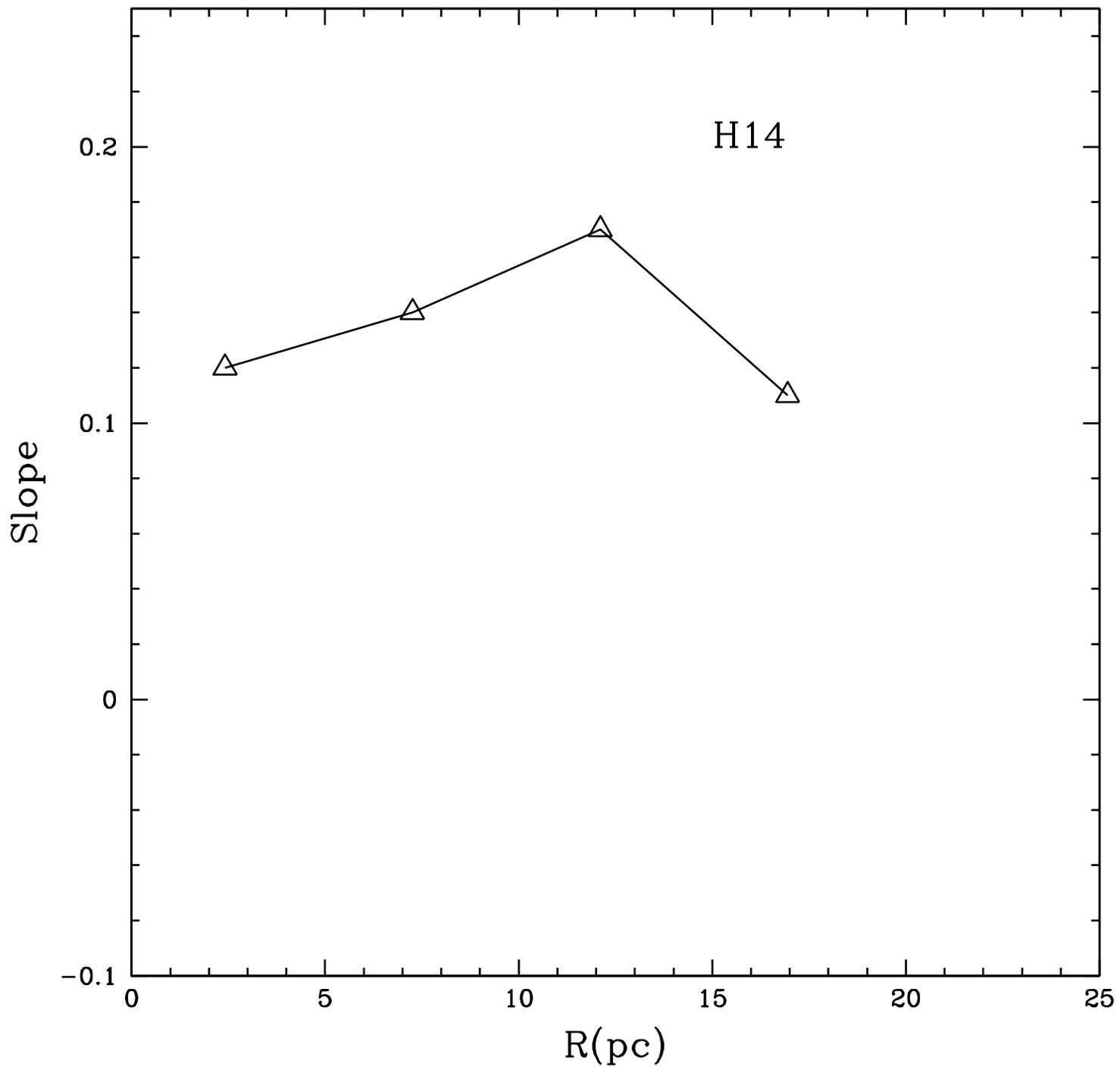
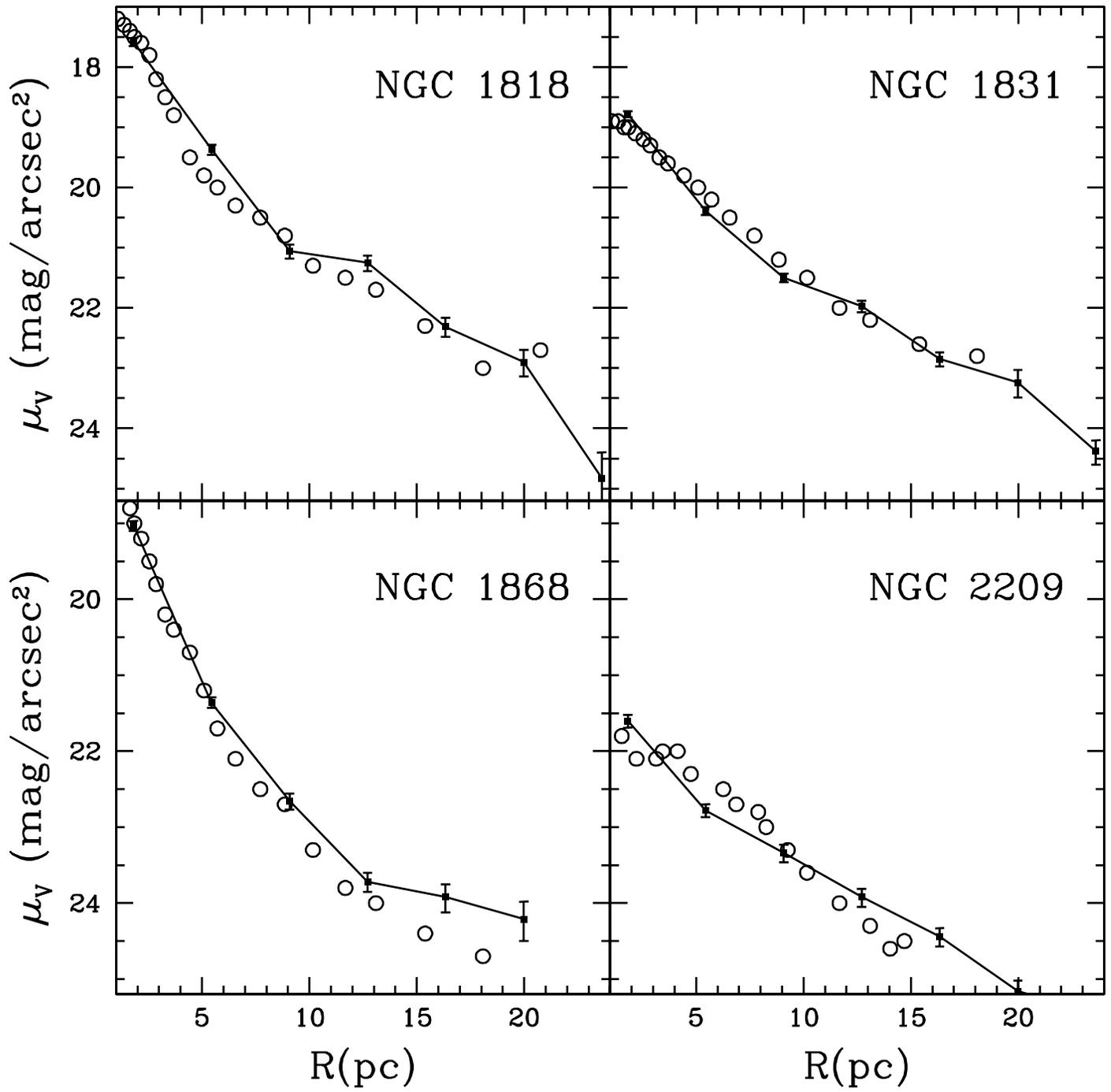
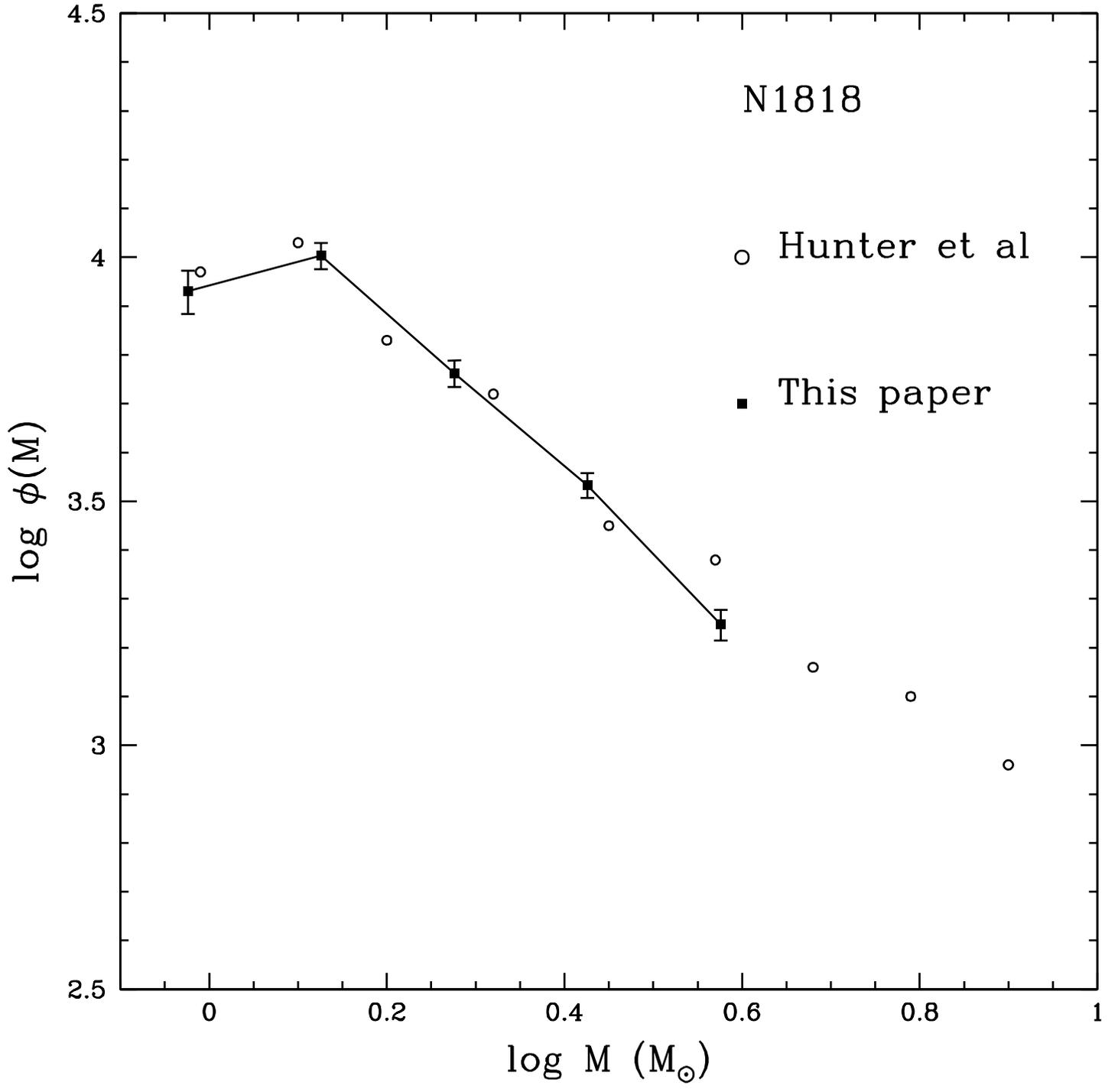
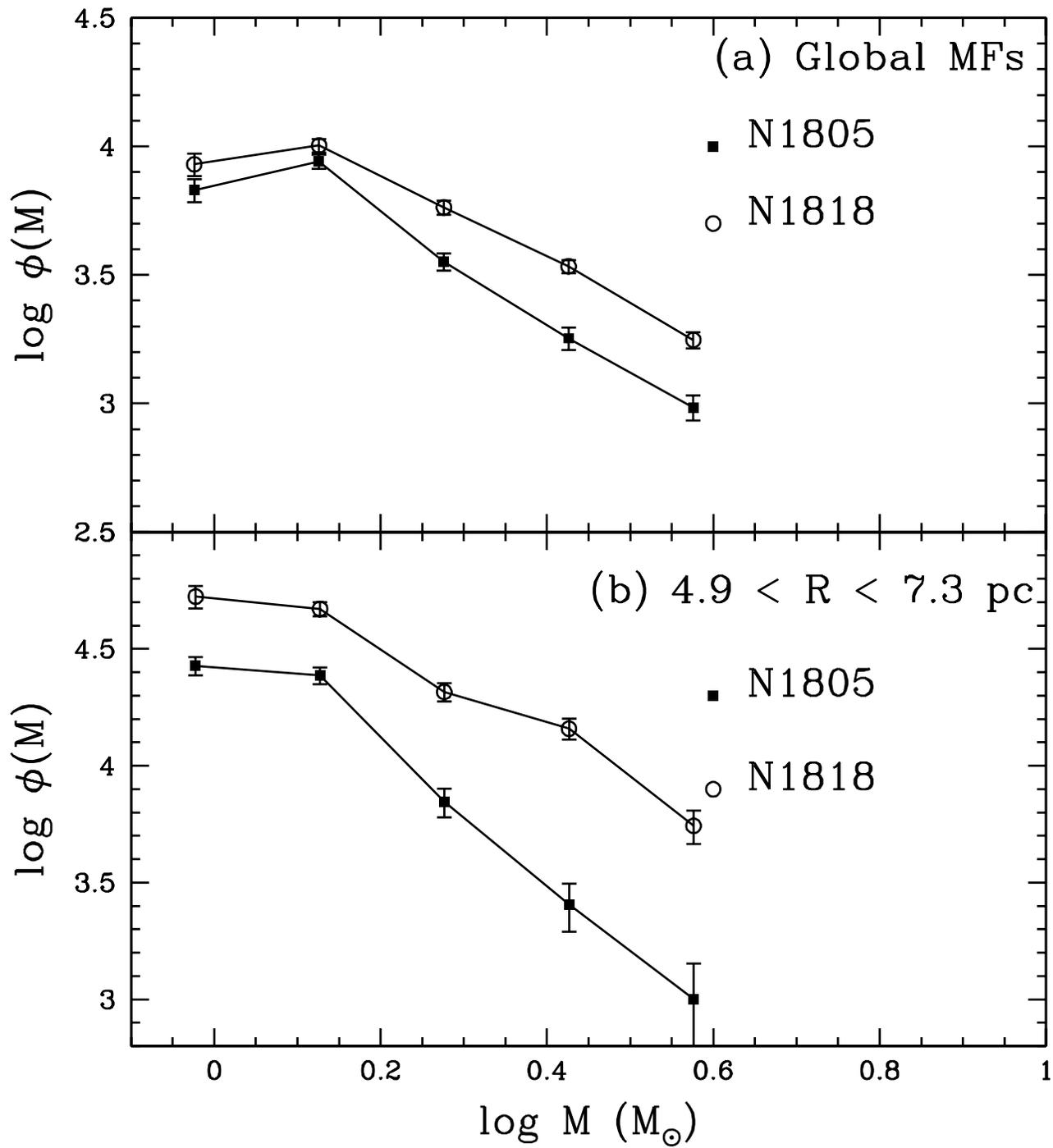
\end{document}